\documentclass[aps,superscriptaddress,a4paper,11pt,showpacs]{revtex4}
\usepackage[colorlinks=true, pdfstartview=FitV, linkcolor=blue, citecolor=red, urlcolor=black]{hyperref}
\usepackage[latin1]{inputenc}
\usepackage{amsmath}
\usepackage{amsfonts}
\usepackage{amssymb}
\usepackage{graphicx}

\newcommand{\be}{\begin{equation}}
\newcommand{\ee}{\end{equation}}
\newcommand{\ben}{\begin{eqnarray}}
\newcommand{\een}{\end{eqnarray}}
\newcommand{\bes}{\begin{subequations}}
\newcommand{\ees}{\end{subequations}}

\newcommand{\bfi}{\begin{figure}}
\newcommand{\efi}{\end{figure}}
\newcommand{\bc}{\begin{center}}
\newcommand{\ec}{\end{center}}
\newcommand{\sech}{\mbox{sech}}
\newcommand{\arcsinh}{\mbox{arcsinh}}
\newcommand{\LL}{{\cal L}}

\newcommand{\arctanh}{\mbox{arctanh}}

\begin{document}

\title{Cuscuton kinks and branes}
\author{I. Andrade}\email{andradesigor0@gmail.com}
\affiliation{Departamento de F\'\i sica, Universidade Federal da Para\'\i ba, 58051-970 Jo\~ao Pessoa, PB, Brazil}
\author{M.A. Marques}\email{mam.matheus@gmail.com}\affiliation{Departamento de F\'\i sica, Universidade Federal da Para\'\i ba, 58051-970 Jo\~ao Pessoa, PB, Brazil}
\author{R. Menezes}\email{rmenezes@dce.ufpb.br}
\affiliation{Departamento de Ci\^encias Exatas, Universidade Federal
da Para\'{\i}ba, 58297-000 Rio Tinto, PB, Brazil}
\affiliation{Departamento de F\'\i sica, Universidade Federal da Para\'\i ba, 58051-970 Jo\~ao Pessoa, PB, Brazil}
\begin{abstract}
In this paper, we study a peculiar model for the scalar field. We add the cuscuton term in a standard model and investigate how this inclusion modifies the usual behavior of kinks. We find the first order equations and calculate the energy density and the total energy of the system. Also, we investigate the linear stability of the model, which is governed by a Sturm-Liouville eigenvalue equation that can be transformed in an equation of the Shcr\"odinger type. The model is also investigated in the braneworld scenario, where a first order formalism is also obtained and the linear stability is investigated. Surprisingly, the solutions are independent from the cuscuton term. We also show that this modification allows for the presence of analytical results in the variables that describe the transformed Schr\"odinger-like equation.
\end{abstract}
\maketitle

\section{Introduction}
In physics, topological structures appear in a diversity of contexts, as in the structure formation in the primordial Universe, in the cosmic evolution and in other topics in high energy physics \cite{vilenkin,manton,weinberg}. They arise in one, two and three spatial dimensions as kinks, vortices and monopoles, respectively. Kinks are the simplest ones and appear as static solutions of the equations of motion in an action of a single real scalar field \cite{vachaspati}. Due to their simplicity, they find several applications in condensed matter and can be used to describe some specific behaviors in superconductors, magnetic materials and topological insulators, for instance; see Ref.~\cite{fradkin}.

Topological structures may also be investigated in non canonical models. The motivation comes from the context of inflation \cite{cosm1}, in which generalized scalar kinetic terms may drive the inflationary evolution without the presence of a potential. Generalized models were also used as a tentative solution to explain why the universe is in an accelerated expansion in a late evolution stage \cite{cosm2,cosm3}. Regarding topological objects, non canonical models can be used to investigate new behavior of vortices, as to simulate Chern-Simons properties \cite{jackiwcs} using a generalized Maxwell-Higgs model \cite{G1,G2}, and to describe their compactification \cite{compvortex,compcs,godvortex}, which may map the magnetic field of a infinitely long solenoid.

 Kinklike structures may also be found in generalized models \cite{stabbabichev,trilogy1,trilogy2}. The non canonical actions allows for the presence of new features. For instance, in Refs.~\cite{trodden,twinb1,twinb2,twinb3,twinb4,twinb5,twinb6}, it was presented non canonical models that share the same field profile and energy density, which are called twinlike models. Other investigations with generalized models deal with models similar to the Born-Infeld type \cite{bil}, models that share the same energy density and linear stability \cite{fphil}, tachyon dynamics and condensation \cite{S,S0,S1,T1,T2,D,T3} and tachyonic dark energy in the context of cosmology \cite{pad,baz,berto}; see also Refs.~\cite{gen1,gen2,gen3,gen4,gen5,gen6,gen7}.
 
Among the generalized models, a special one was introduced in Refs.~\cite{cuscuton1,cuscuton2}, where one started to consider the cuscuton modification of gravity. It consists in a class of actions with a non-canonical kinetic term that lead to an effective violation of the null energy condition (NEC) in cosmological backgrounds with the matter sources satisfying the NEC. In this case, in the cosmological homogeneous limit for the field, the equation of motion does not present time dependent second order contributions and the field becomes non dynamical. As we show in this paper, the cuscuton term, when added to the canonical kinematics in the Lagrangian density of a single real scalar field, preserves the standard form of the equation of motion for static configurations, i.e., the solution of the modified model is the same of the canonical one. However, the other properties of the system such as the energy density and the linear stability are affected by the new term. Over the years, several papers that deal with the cuscuton term have appeared; see Refs.~\cite{cuscuton3,cuscuton4,cuscuton5,cuscuton6,cuscuton7,cuscuton8}.

An interesting application of kinklike structures appears in the braneworld scenario \cite{RS,GW,MC,brane,cvetic,gremm}. In this context, they may be used to model the single extra dimension of an infinite extent, which gives rise to the concept of thick brane. This feature allows for the presence of the so called hybrid branes, which are modeled by compact scalar field profiles \cite{fktc,fkthc}. It is interesting to highlight that this modelling can also be done with the use of generalized models, that support the presence of a first order framework and stable branes \cite{trilogy3,bil}. Twinlike models may also be found in this scenario \cite{twinb3}.

In this work, we investigate the presence of kinks in a generalized model that adds the cuscuton term to the standard kinematics of the scalar field. In Sec.~\ref{sec2}, we describe the model in the flat spacetime and its properties. We also investigate the linear stability of the model and show that, even though it is modified by the new term, the kink is stable. This is achieved by transforming the Sturm-Liouville equation into a Schr\"odinger-like equation. Usually, this cannot be done analytically. To illustrate our model, we present two examples, one of them being completely described by analytical quantities. In Sec.~\ref{sec3}, we extend the model to the braneworld scenario, also calculating how the cuscuton term modifies the equations of motion, energy density and stability equation. We show that the brane is stable and provide an example that also admits analytical expressions in the study of stability. We end our paper in Sec.~\ref{sec4} with our conclusions and perspectives.

\section{The Model}\label{sec2}
We start by considering a class of models described by the action for a single real scalar field $\phi$, with the Lagrangian density
\be\label{lmodel}
{\cal L} = X +  \frac{2f(\phi)X}{\sqrt{|2X|}} - V(\phi),
\ee
where the dynamical term of the scalar field is $X=\frac12 \partial_\mu \phi \partial^\mu \phi$ and the potential is described by $V(\phi)$. We work in Minkowski spacetime, with metric tensor $\eta_{\mu\nu}=\rm{diag}(+,-)$. Furthermore, we take dimensionless fields and spacetime coordinates. \textit{A priori}, $V(\phi)$ and $f(\phi)$ are arbitrary functions of $\phi$, which identify how the scalar field self-interacts. Note that $f(\phi)=0$ recovers the standard case $\LL=X-V(\phi)$, that was studied exhaustively in Refs.~\cite{vachaspati,manton}. The second term in the above Lagrangian density comes from Ref.~\cite{cuscuton1}. Without the first term, $X$, associated to the standard kinematics of the scalar field, the function $f(\phi)$ may be set to a constant with no loss of generality; this can be achieved by making a field redefinition. In the present case, this is not valid anymore; one may perform the change $\phi\to \int d\phi/f(\phi)$ to show that Eq.~\eqref{lmodel} becomes
\be
\LL=\frac{1}{f^2}X + \frac{2X}{\sqrt{|2X|}} - \tilde{V}(\phi),\quad\text{where}\quad \tilde{V}(\phi)=V\left(\int \frac{d\phi}{f(\phi)}\right).
\ee
Thus, it is not possible to eliminate the function $f$ through a field redefinition.

By varying the action with respect to $\phi$, one can show that the equation of motion is given by
\be
\partial_\mu  \left(\left(1\!+\! \frac{f(\phi)}{\sqrt{|2X|}}\right)\partial^\mu \phi \right)=\frac{2f_\phi X}{\sqrt{|2X|}} - V_\phi,
\ee 
where $V_\phi=dV/d\phi$ and $f_\phi=df/d\phi$. The above equation can be expanded in the form
\ben\label{eom}
\left[\!\left(1\!+\! \frac{f(\phi)}{\sqrt{|2X|}}\right) \eta^{\mu\nu} \!-\!\frac{f(\phi)\partial^\mu \phi \partial^\nu \phi}{2X\sqrt{|2X|}}\right]\!\partial_\mu \partial_\nu \phi + V_\phi=0.
\een
The invariance over spacetime translations leads to the following energy-momentum tensor
\be\label{tmunu}
T_{\mu\nu}= \left(1+ \frac{f(\phi)}{\sqrt{|2X|}}\right)\partial_\mu \phi \partial_\nu \phi -\eta_{\mu\nu}\LL.
\ee
It is conserved, $\partial_\mu T^{\mu\nu}=0$, for field configurations that obey the equation of motion in Eq.~\eqref{eom}. Following Ref.~\cite{trilogy1}, we investigate the Null Energy Condition (NEC), which imposes $T_{\mu\nu}n^\mu n^\nu \geq 0$, where $n^\mu$ is a null vector ($\eta_{\mu\nu}n^\mu n^\nu =0$). This leads to 
\be\label{NEC}
1 + \frac{f(\phi)}{\sqrt{|2X|}}\geq 0.
\ee

For static configurations, $\phi=\phi(x)$, the equation of motion drastically
simplifies to
\be\label{eomstatic}
\phi^{\prime\prime} = V_\phi,
\ee
where $\phi$ connects two neighbor minima of the potential, which we call $v_\pm$. Here, the prime stands for the derivative with respect to $x$. It is interesting to note that this solutions do not depend on $f(\phi)$, i.e., for a given potential $V(\phi)$, we take the same set of solutions for any $f(\phi)$, including the standard case, $f(\phi)=0$. In this case, the energy and stress densities can be calculated from Eq.~\eqref{tmunu}, that leads to
\ben
T_{00}&=& \frac12 \phi^{\prime2} + {f(\phi)}{|\phi^\prime|}+V(\phi),  \\ \label{stress}
T_{11}&=&\frac12 \phi^{\prime2}-V(\phi), 
\een
respectively. By integrating $T_{00}$ all over the space, we get the energy
\be
E=\int^\infty_{-\infty}\,dx\left(\frac12 \phi^{\prime2} + {f(\phi)}{|\phi^\prime|}+V(\phi)\right).
\ee

The equation of motion \eqref{eomstatic} is of second order and usually presents nonlinearities engendered by the potential. Then, it is very hard to calculate its solutions. In order to simplify the problem, we integrate it to get
\be
\frac12 \phi^{\prime2}-V(\phi)=C.
\ee 
By comparing it with the stress density in Eq.~\eqref{stress}, we can identify $T_{11}=C$. As shown in Ref.~\cite{trilogy1}, the Derrick's theorem extension affirms that only stressless solutions ($C=0$) can be stable. Thus, we write
\be\label{eqfirstC0}
\phi^{\prime2}=2V(\phi).
\ee
The above equation admits two signs for the derivative of the field. Nonetheless, they are related by the change $x\to-x$. Thus, from now on, we only consider 
\be\label{fo}
\phi^\prime = \sqrt{2V(\phi)},
\ee
whose solution we call kink. As in Eq.~\eqref{eomstatic}, this equation does not depend on $f(\phi)$. Despite that, the energy density of the solutions that obey Eq.~\eqref{eqfirstC0} depends explicitly on $f(\phi)$  
\be\label{rho}
\begin{split}
	T_{00}&=2V(\phi)+f(\phi)\sqrt{2V}\\
	      &=(\sqrt{2V}+f(\phi))\phi^\prime.
\end{split}
\ee
Following the formalism introduced in Ref.~\cite{trilogy2}, we introduce an additional function to define 	
\be\label{Wp}
W_\phi=\sqrt{2V}+f(\phi).
\ee
Substituting this in Eq.~\eqref{rho}, we get
\be\label{ew}
E=W(\phi(v_1))-W(\phi(v_2)),
\ee
which shows that it is possible to calculate the energy without knowing the explicit solutions. For functions $f(\phi)$ that obey the NEC \eqref{NEC}, we get $W_\phi>0$, therefore the $W$ function is monotonic in the domain of the solutions of Eq.~\eqref{eqfirstC0}.  

Now we turn our attention to investigate the linear stability of the solution. We take $\phi(x,t)=\phi(x)+\eta(x,t)$, supposing that $\eta(x,t)$ is a small fluctuation around the static solution $\phi(x)$. In this case we get, going up first-order in the fluctuations, the following equation for $\eta$
\be\label{stabeqt}
A^{-2}\ddot\eta-\eta^{\prime\prime} +V_{\phi\phi}\eta=0,
\ee
where $V_{\phi\phi} = d^2V/d\phi^2$ and
\be\label{A}
A^2= \left(1\!+\! \frac{f(\phi)}{\sqrt{2V}}\right)^{-1}.
\ee
We impose $A^2\geq 0$ in this equation to preserve its hyperbolicity, as in Ref.~\cite{stabbabichev}. The stability equation \eqref{stabeqt} admits the variables $x$ and $t$ to be separated as
\be
\eta(x,t)=\sum_i \eta_i(x)\cos(\omega_i t).
\ee
In this case, we obtain
\be\label{stabeq}
-\eta_i^{\prime\prime} +V_{\phi\phi}\eta_i = A^{-2} \omega^2\eta_i.
\ee
This is a Sturm-Lioville (SL) eigenvalue equation. Only for the standard case ($f(\phi)=0$), it becomes the Schr\"odinger equation. In general the SL equation supports scattering states, bound states and, possibly, half-bound states. For homogeneous solutions $\phi=v_\pm$, this equation is simply the Laplace/Helmholtz equation. One may show the zero mode is $\eta_0\propto\phi^\prime$, which is an evidence of the translational invariance of the solution.

The SL equation \eqref{stabeq} may be transformed into a Schr\"odinger-like equation with the following change of variables
\bes
\ben\label{dxdz}
dz &=& A^{-1} dx,\\ \label{ueta}
u_i &=& A^{-\frac12}\eta_i,
\een
\ees
which leads to the stability equation 
\be\label{stabu}
-{u_i}_{zz}+U(z)u_i=\omega^2 u_i,
\ee
where the stability potential is written as
\be\label{potestschr}
U(z)=\sqrt{\frac{A}{2V}} \left(\sqrt{\frac{2V}{A}}\right)_{zz}.
\ee
In order to write the correspondence between $x$ and $z$, one has to integrate Eq.~\eqref{dxdz} and find $x$ as a function of $z$. This task is usually very complicated. Thus, the explicit form of the stability potential and eigenfunctions are commonly unknown. In Refs.~\cite{trilogy1,trilogy2}, one may find how this procedure works for models that lead to a constant $A$. Here, we introduce new models with non constant $A$ that supports this correspondence, which is an unprecedented feature in the study of the stability of kinklike solutions.

The stability equation \eqref{stabu} can be factorized in the form
\be\label{stabfac}
S^\dagger S u_i = \omega^2 u_i,
\ee
where the operator $S$ is given by
\be\label{ssup}
S= \frac{d}{dz} + g(z), \quad\text{with}\quad g(z) = \left[\ln\left(\sqrt{\frac{2V}{A}}\right)\right]_z.
\ee
This procedure makes possible to write the stability potential as $U(z) = g^2 + g_z$. The factorized equation \eqref{stabfac} may be multiplied by $u_i$ and integrated to show that 
\be
\omega_i^2 = \frac{\int^\infty_{-\infty} dz \left|S u_i \right|^2}{\int^\infty_{-\infty} dz \left|u_i \right|^2}.
\ee
This means that $\omega_i^2\geq0$. Since there is no negative eigenvalue, the kinklike solutions of Eq.~\eqref{eqfirstC0} are linearly stable. The interest of this paper is to investigate the modifications introduced by $f(\phi)$. Below, we consider two distinct examples of $f(\phi)$ for both the $\phi^4$ and the sine-Gordon potential.

\subsection{Polynomial potential}\label{poly}
Our model is defined by the function $f(\phi)$ and the potential $V(\phi)$. Before suggesting the form of $f(\phi)$, we take the well-known $\phi^4$ potential
\be\label{potential_ex}
V(\phi)=\frac12 (1-\phi^2)^2.
\ee  
It presents a $Z_2$ symmetry and has two minima located at $\phi_\pm=\pm 1$ and a maximum at $\phi_0=0$, with $V(\phi_0)=1/2$. In this case, the kink solution of Eq.~\eqref{fo} is
\be\label{tanh}
\phi(x)=\tanh(x), 
\ee
regardless the $f(\phi)$ that is chosen. In the standard case, $f(\phi)=0$, the energy density is
\be 
T_{00}=2V(\phi)=S^4(x),
\ee
with $S(x)=\sech(x)$ and energy $E=4/3$. The stability potential is given by
\be
U(x) = 4-6\,S^2(x), 
\ee
which is a modified P\"oschl-Teller potential whose eigenvalues are $\omega^2=0$ and $\omega^2=3$ for the bound states. Below, we consider two examples of functions $f(\phi)$ that modifies the model with the $\phi^4$ potential.

\subsubsection{Example 1}\label{ex1}
Firstly, we study the simplest case, which arises for $f(\phi)=f_0$, where $f_0$ is a positive parameter to satisfy the NEC in Eq.~\eqref{NEC}. We can write
\be
W(\phi)=(1+f_0)\phi-\frac13\phi^3.
\ee 
The energy density \eqref{rho} is
\be 
T_{00}=(1-\phi^2)(1-\phi^2+f_0)=S^2(x)(f_0+S^2(x)).
\ee
We note that $T_{00}(x=0)=1+f_0$. The energy can be calculated from Eq.~\eqref{ew}, which leads to $E=4/3+2f_0$.

In order to study the stability, we calculate $A$ from Eq.~\eqref{A}
\be
A^2 = \frac{1}{1+f_0\cosh^2(x)},
\ee
which is positive, obeying the hyperbolic condition. From Eq.~\eqref{dxdz} we may find $z$ as a function of $x$. However, the expression appears in terms of elliptic functions, whose inverse function cannot be found explicitly. Even so, a numerical approach may be used, but it is not our purpose here.

\subsubsection{Example 2}\label{ex2}
\begin{figure}[t!]
\centering
\includegraphics[width=6cm]{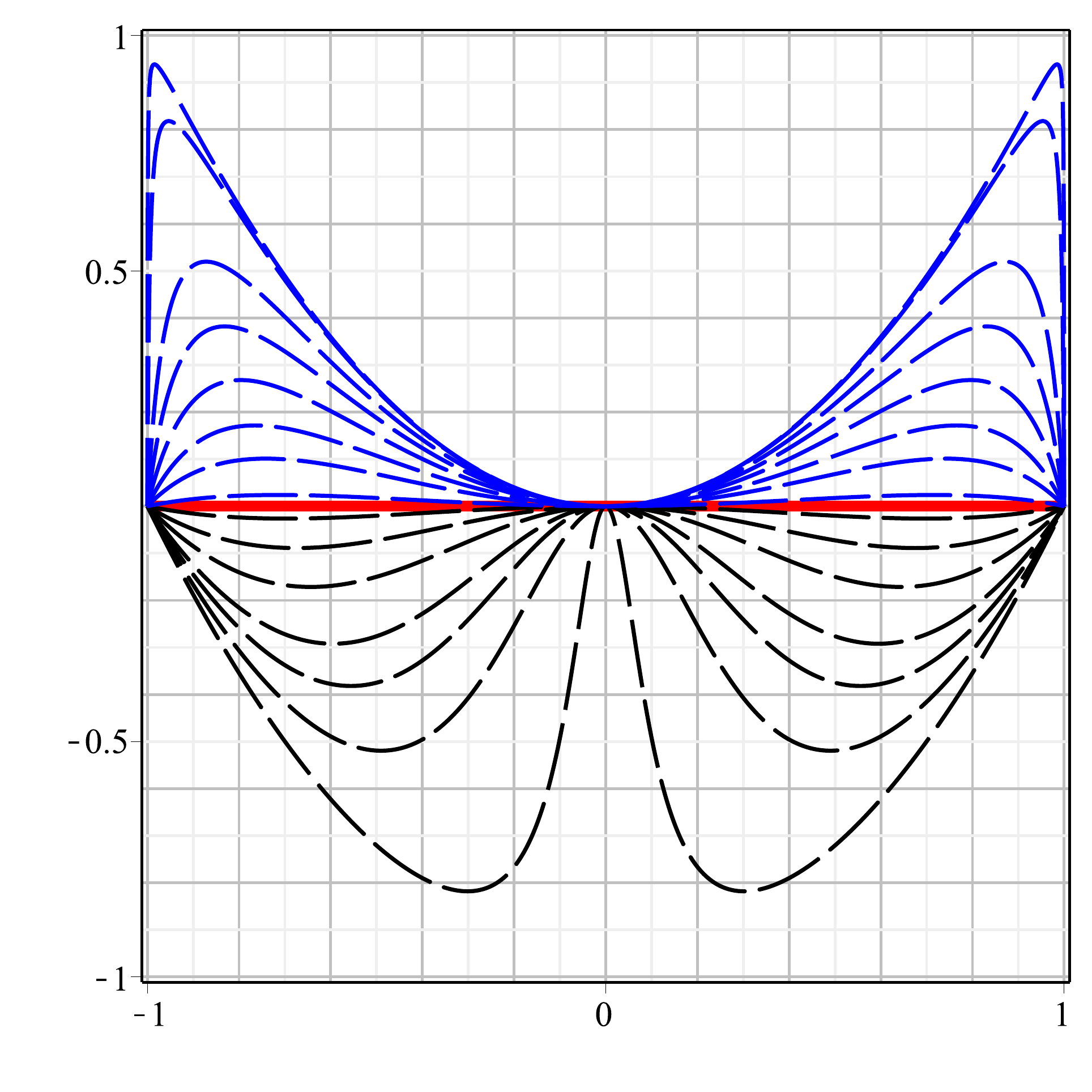}
\caption{The function $f(\phi)$ in Eq.~\eqref{f_a_sinh} for $0<a\leq1$ (black) and for $a>1$ (blue) for several values of $a$. The red line represents the limit $a=1$ that is the standard case.}
\label{figf2}
\end{figure}

\begin{figure}[!htb]
\centering
\includegraphics[height=5cm]{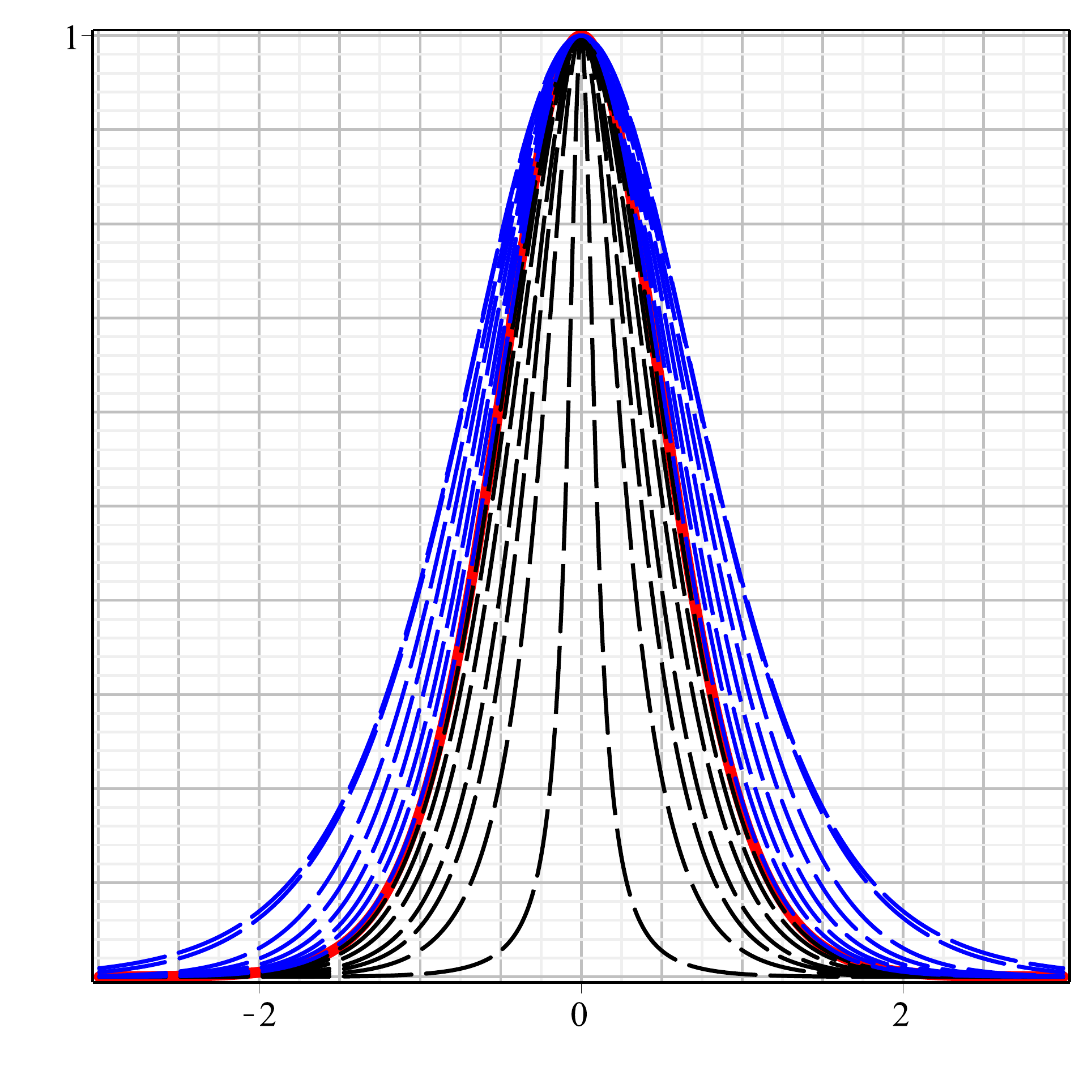}
\includegraphics[height=5cm]{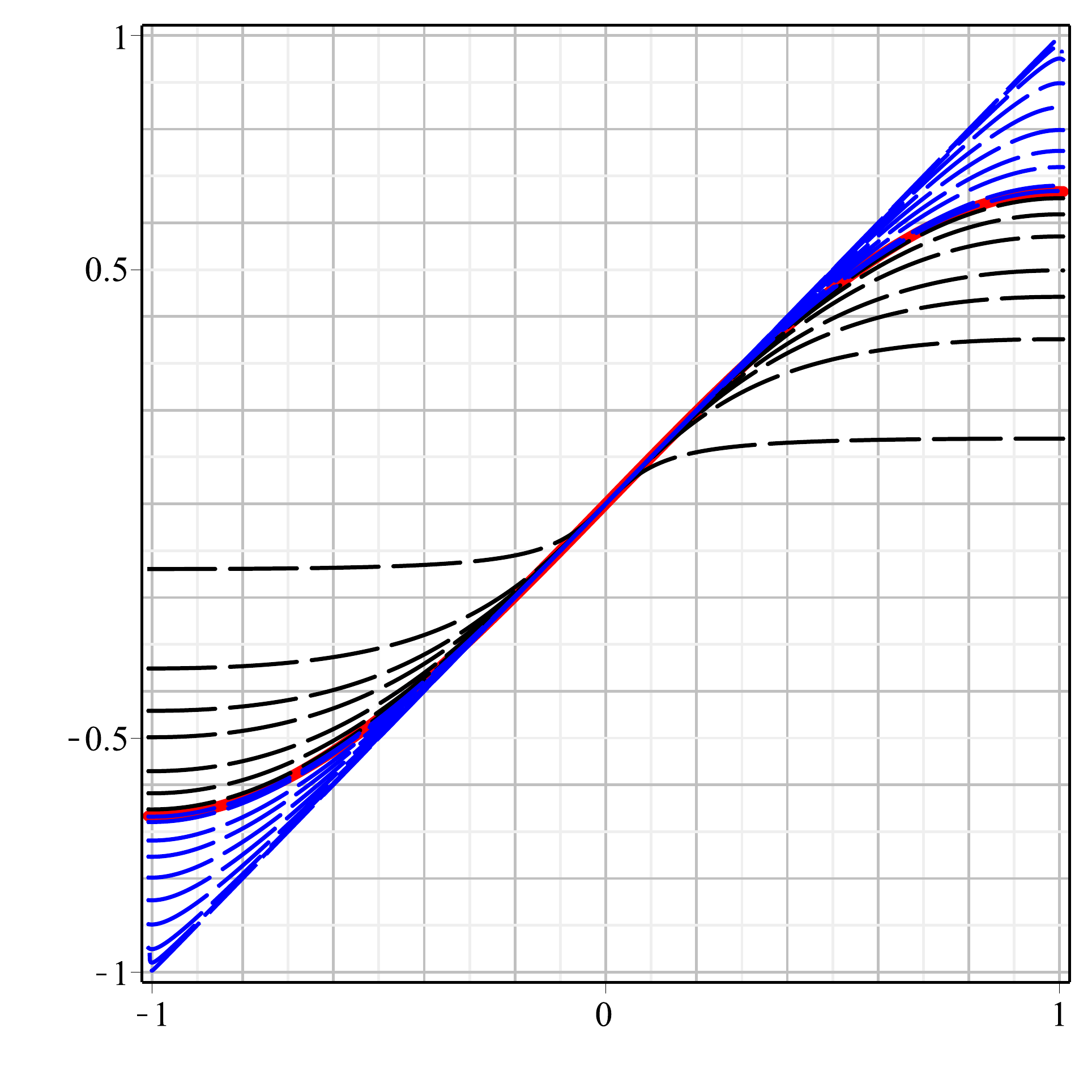}
\caption{The energy density as a function of $x$ in \eqref{rho2} (left) and the function $W(\phi)$ in Eq.~\eqref{W2} (right) for several values of $a$. The colors of the lines follow the previous figure.}
\label{figw2}
\end{figure}

Our second model arises with a function that depends on the parameter $a$.
\be\label{f_a_sinh}
f(\phi)=\frac{(a-1)\phi^2 (1-\phi^2)}{a-(a-1)\phi^2}.
\ee
In order to find the range in which the parameter $a$ exists, we use the NEC in Eq.~\eqref{NEC}
\be
\frac{a}{a-(1-a)\phi^2}\geq 0.
\ee
which is only valid for $a\geq0$. As we will show further, we do not consider $a=0$ because it leads to null energy density. The case $a=1$ is such that $f_1(\phi)=0$ and the standard case is recovered. For $a\to\infty$, we have $f_\infty(\phi)=\phi^2$. In Fig.~\ref{figf2}, we plot the function $f(\phi)$ for several values of $a$. 
%%%%%%%%%%%%%%%%%%%%
%%%%%%%%%%%%%%%%%%%
The energy density is
\be\label{rho2}
T_{00}=\frac{a(1-\phi^2)^2}{a-(a-1)\phi^2}=\frac{a\,S^4(x)}{1+(a-1)\,S^2(x)}.
\ee	
For $a=1$, we have $T_{00}=S^4(x)$. In the limit $a\to\infty$, $T_{00}=S^2(x)$. In Fig.~\ref{figw2}, we see the energy density for several values of $a$.
%%%%%%%%%%%%%%%%%%%
%%%%%%%%%%%%%%%%%%%%
From Eq.~\eqref{Wp}, we can write $W(\phi)$ as  
\be\label{W2}
W(\phi)=\frac{a\,\phi}{a-1} +\frac{\sqrt{a}\,M(\phi)}{|a-1|^{\frac32}},
 \ee
with
\be
M(\phi)=
\begin{cases}
\displaystyle\arctan\left(\sqrt{\frac{1-a}{a}}\,\phi\right)
,&a\leq1\\
-\displaystyle\arctanh\left(\sqrt{\frac{a-1}{a}}\,\phi\right), & a>1.
\end{cases}
\ee
We can see that $W(\phi)=\phi-\frac13 \phi^3$ for $a=1$ and $W(\phi)=\phi$ for $a\to\infty$. The energy can be evaluated through Eq.~\eqref{ew}. For $a\leq1$, the energy is $E=2a/(a-1) + 2\arctan(\sqrt{(1-a)/a})\,\sqrt{a}/|a-1|^{3/2}$ and for $a>1$, the energy is $E=2a/(a-1) - 2\,\arctanh(\sqrt{(a-1)/a})\,\sqrt{a}/|a-1|^{3/2}$. The energy, then, is controlled by the parameter $a$. In the limit $a\to1$, $E\to4/3$ and for $a\to\infty$, $E\to2$.

\begin{figure}[t!]
\centering
\includegraphics[width=5cm]{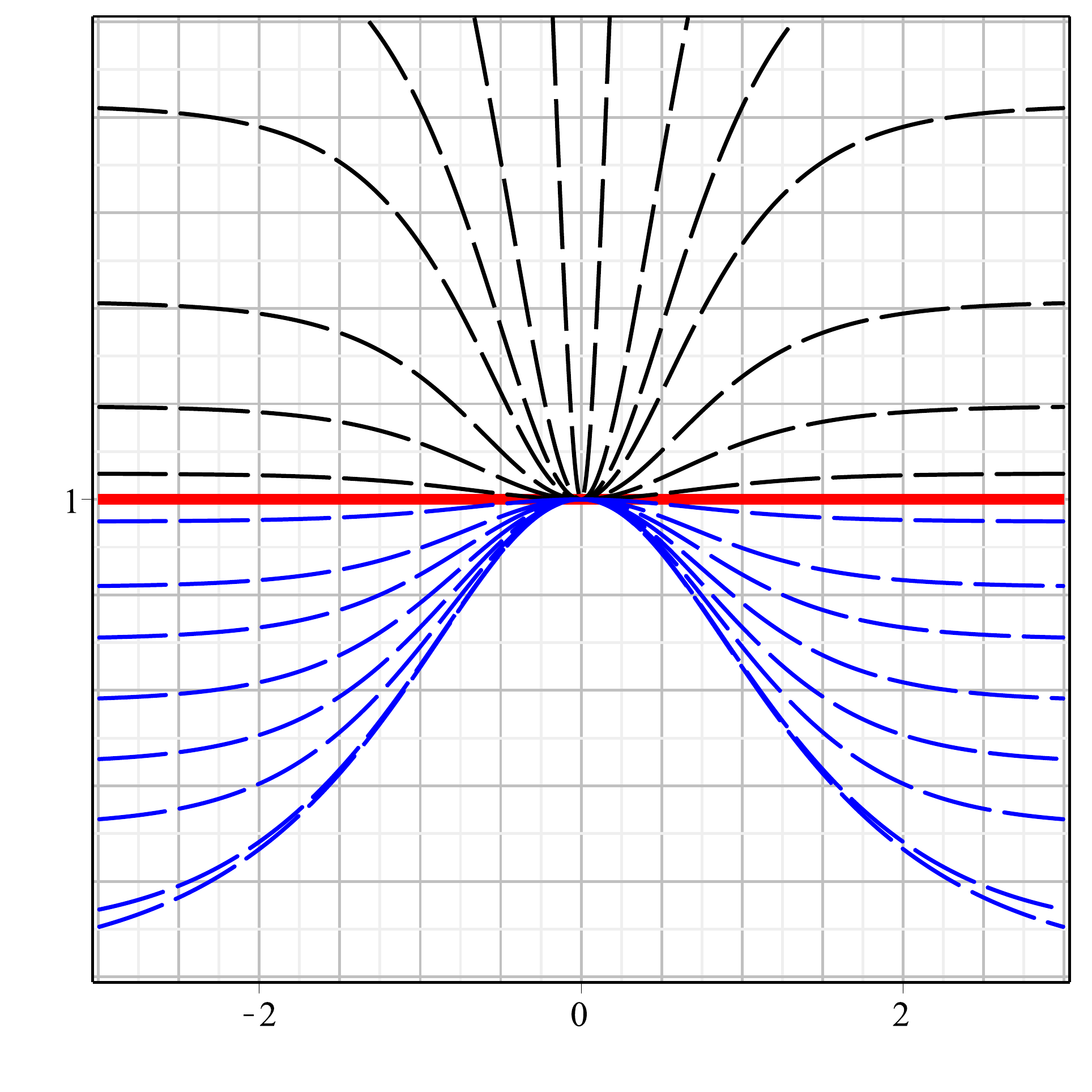}
\includegraphics[width=5cm]{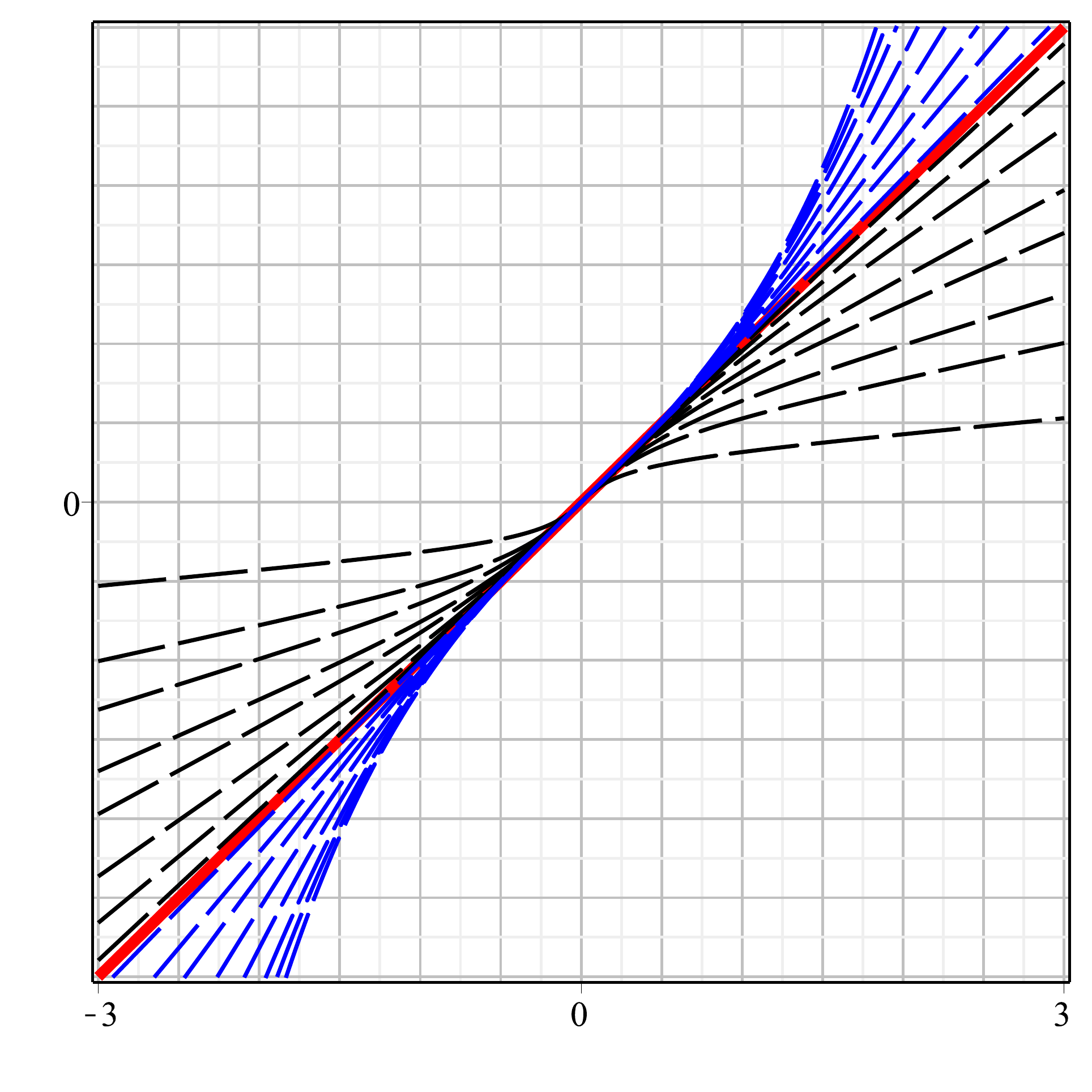}
\caption{The function $A(x)$ in Eq.~\eqref{A2} (left) and the rescaled coordinate $z$ as a function of $x$ in Eq.~\eqref{zx2} (right).}
\label{figa2}
\end{figure}

\begin{figure}[htb!]
\centering
\includegraphics[width=5cm]{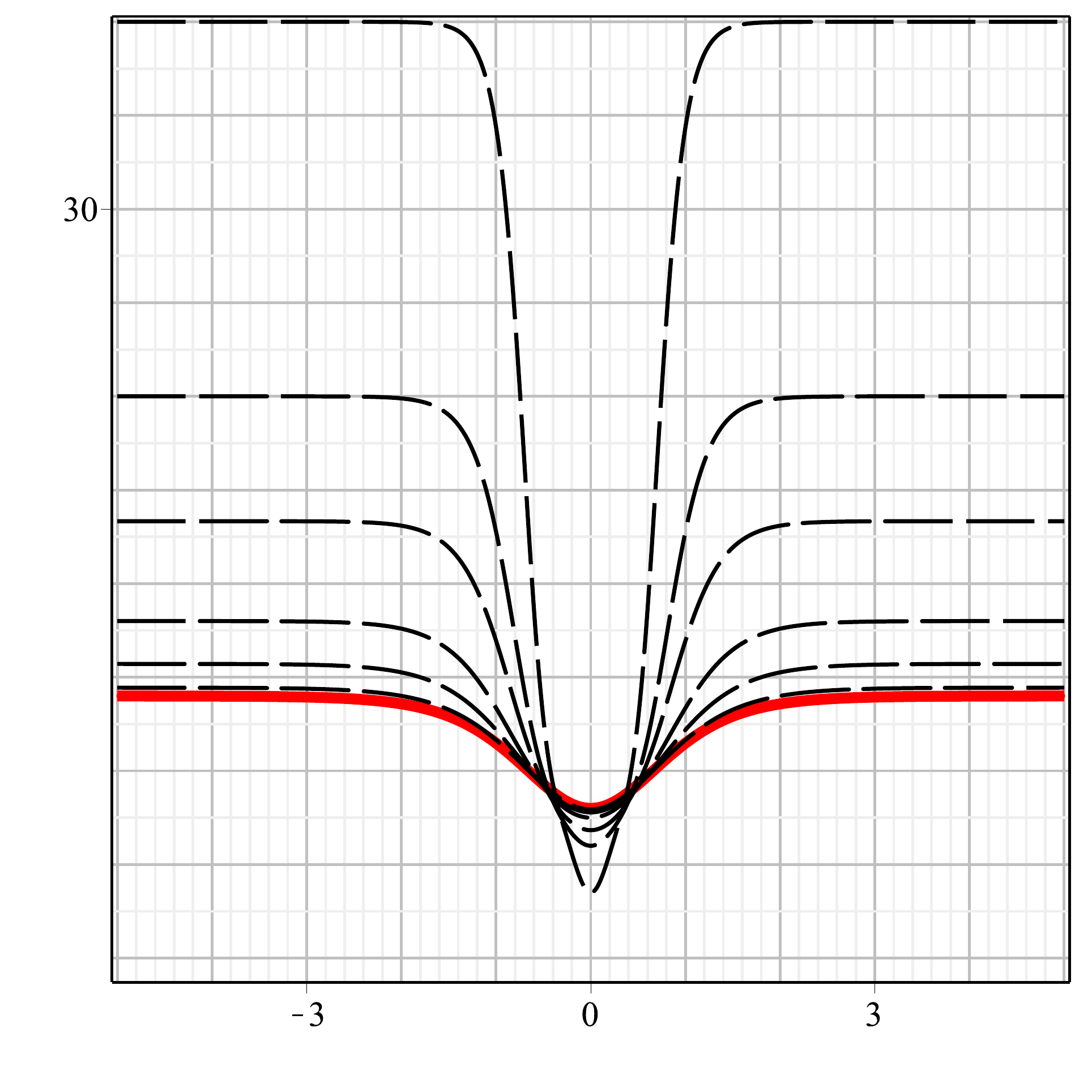}
\includegraphics[width=5cm]{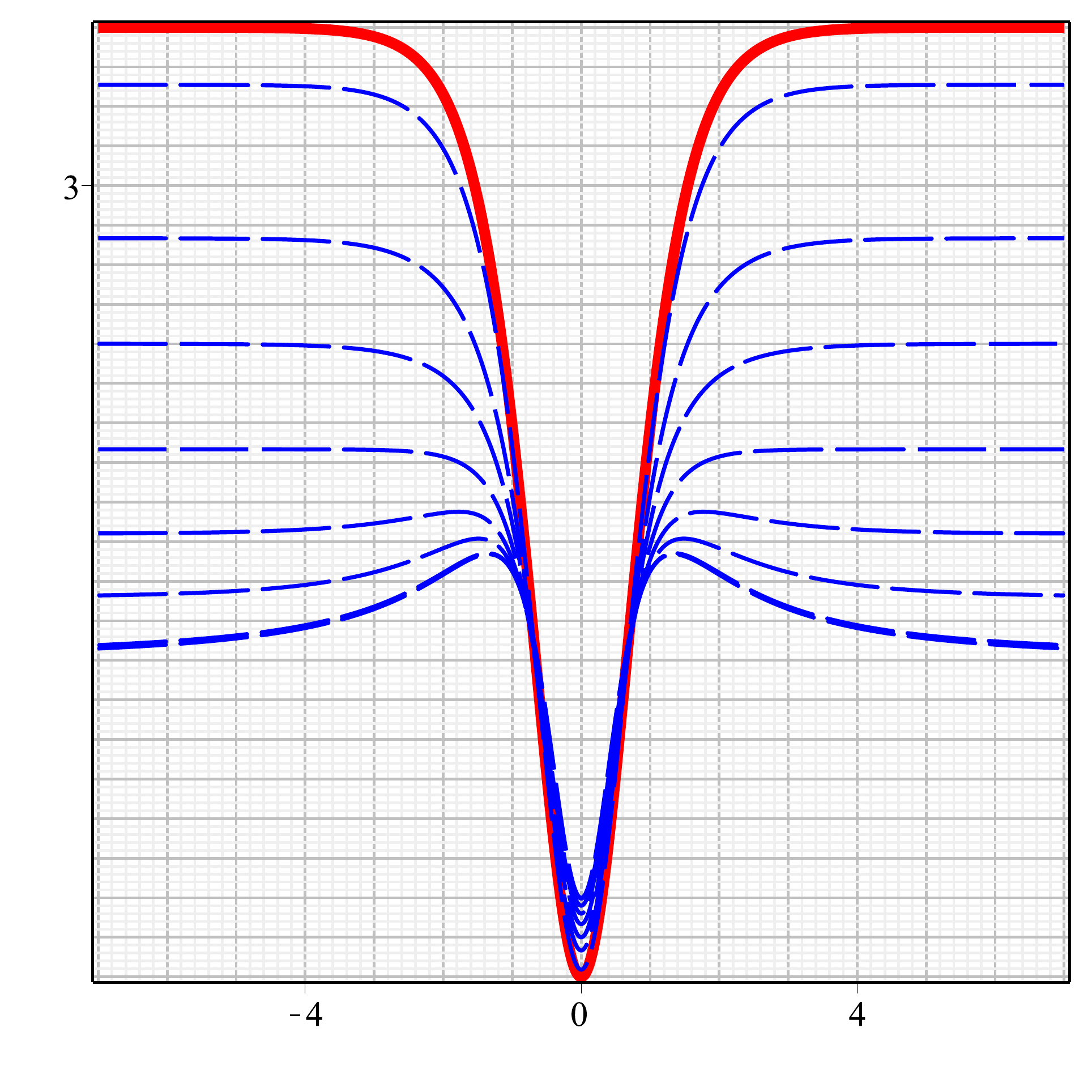}
\caption{The stability potential $U(z)$ in Eq.~\eqref{Uest1} for $0<a\leq1$ (black) and for $a>1$ (blue) for several values of $a$. The red line represents the limit $a=1$ that is the standard case.}
\label{figu}
\end{figure}

The function $A$ in Eq.~\eqref{A} takes the form 
\be\label{A2}
A^2=1-\frac{a-1}{a}\phi^2=\frac1a+\frac{a-1}{a}S^2(x),
\ee
which is always positive, preserving the hyperbolicity of the stability equation. We have $A^2=1$ for $a=1$ and $A^2 = S^2(x)$ for $a\to\infty$, as shown in Fig.~\ref{figa2}, in which the function $A$ is plotted for several values of $a$.
To study the stability, we make the change of variables proposed in Eq.~\eqref{dxdz}:
\be\label{zx2}
z = \sqrt{a}\,\arcsinh\left(\frac{\sinh(x)}{\sqrt{a}}\right).
\ee
The above equation was the result of an integration. In this model, we can go further and find $x$ as a function of $z$ by inverting the above equation
\be\label{xzana}
x=\arcsinh\left(\sqrt{a}\sinh\left(\frac{z}{\sqrt{a}}\right)\right).
\ee
We also make a change in the fluctuations, following Eq.~\eqref{ueta}
\be
u = \left(\frac{a}{1+(a-1)S^2(x)}\right)^{\frac14}\eta.
\ee
In Fig.~\ref{figa2}, we show the rescaled coordinate $z$ as a function of $x$. The presence of Eq.~\eqref{xzana} allows us to calculate the explicit form of the function $g(z)$ that appears in the supersymmetric operator in Eq.~\eqref{ssup}. It has the form
\begin{equation}
g(z)=-\frac{1}{2}\tanh(z/\sqrt{a})\frac{\left((a-1)S^2(z/\sqrt{a})-4a\right)}{\sqrt{a}\left((a-1)S^2(z/\sqrt{a})-a\right)}.
\end{equation}
Furthermore, the stability potential in Eq.~\eqref{potestschr} becomes
\begin{equation}\label{Uest1}
	U(z)=-\frac{3(a-1)^2S^6(z/\sqrt{a}) - (7a-1)(a-1)S^4(z/\sqrt{a}) + 4a(5a+1)S^2(z/\sqrt{a}) - 16a^2}{4a\left((a-1)S^2(z/\sqrt{a})-a\right)^{2}}.
\end{equation}
In Fig.~\ref{figu}, we plot the above stability potential for several values of $a$. Here, we emphasize that the presence of the analytical expressions \eqref{zx2}-\eqref{Uest1} is an unprecedented fact and it was only possible due to the specific form of the function $f(\phi)$ in Eq.~\eqref{f_a_sinh} that drives the cuscuton term.
%%%%%%%%%%%%%%%%%%%%
%%%%%%%%%%%%%%%%%%%%
%%%%%%%%%%%%%%%%%%%%%%%%%%%
%%%%%%%%%%%%%%%%%%%%%%%%%%%%%%%%%%%%

\subsection{Sine-Gordon potential}
In Sec.~\ref{poly}, we investigated how the cuscuton term, driven by $f(\phi)$, modifies the $\phi^4$ potential. Here, we consider a non-polynomial potential, the sine-Gordon one, given by
\be\label{potential_sine}
V(\phi)=\frac12 \cos^2(\phi).
\ee
This model is a $\pi$ periodic potential. Its minima are located at $\phi_k=(k-1/2)\pi$ and maxima at $\phi_m=k\pi$ with $k\in\mathbb{Z}$. The potential takes a fixed value in all the maxima: $V(\phi_m)=1/2$. For $k=0$ and $k=1$, we get the minima $\phi=-\pi/2$ and $\phi=\pi/2$, with a maximum at $\phi=0$. We only work in the interval between the aforementioned minima, since the other ones are obtained from the shift $\phi\to\phi+k\pi$. In this case, the solution that lives in this sector is given by
\be\label{sol_sine}
\phi(x)=\arcsin(\tanh(x)).
\ee
For the standard case, $f(\phi)=0$, the energy density is written as
\be 
T_{00}=2V(\phi)=S^2(x),
\ee
with energy $E=2$. The stability potential is
\be
U(x) = 1-2\,S^2(x), 
\ee
which is a modified P\"oschl-Teller potential. Here, however, it is different from the $\phi^4$ potential and admits only the zero mode, with $\omega^2=0$. Next, we present two examples of functions $f(\phi)$ that modifies the model with the sine-Gordon potential.

\subsubsection{Example 1}\label{ex3}
Similarly to the example in Sec.~\ref{ex1}, for the $\phi^4$ potential, we investigate the case $f(\phi)=f_0$ with the sine-Gordon potential in Eq.~\eqref{potential_sine}, with $f_0$ being a positive parameter obeying the NEC in Eq.~\eqref{NEC}. We use Eq.~\eqref{Wp} to get
\be
W(\phi)=\sin(\phi)+f_0\,\phi.
\ee 
The energy density for this model is calculated from Eq.~\eqref{rho}, which leads to 
\be 
T_{00}=\cos(\phi)(\cos(\phi)+f_0)=S(x)(S(x)+f_0).
\ee
The energy density at the origin behaves similarly to the one in $\phi^4$ potential, i.e., $T_{00}(x=0)=1+f_0$. By using Eq.~\eqref{ew} one can show that the energy is $E=2+\pi f_0$. To study the stability, we calculate $A$ from Eq.~\eqref{A}
\be
A^2 = \frac{1}{1+f_0\cosh(x)},
\ee
which obeys the hyperbolic condition. To find the explicit form of $x$ in terms of $z$ from Eq.~\eqref{dxdz} one must use numerical methods, which is not the aim of this paper. So, in the next example we consider another function $f(\phi)$ that allows for the presence of analytical expressions in the study of the linear stability of the model.

\subsubsection{Example 2}\label{ex4}
\begin{figure}[t!]
\centering
\includegraphics[width=6cm]{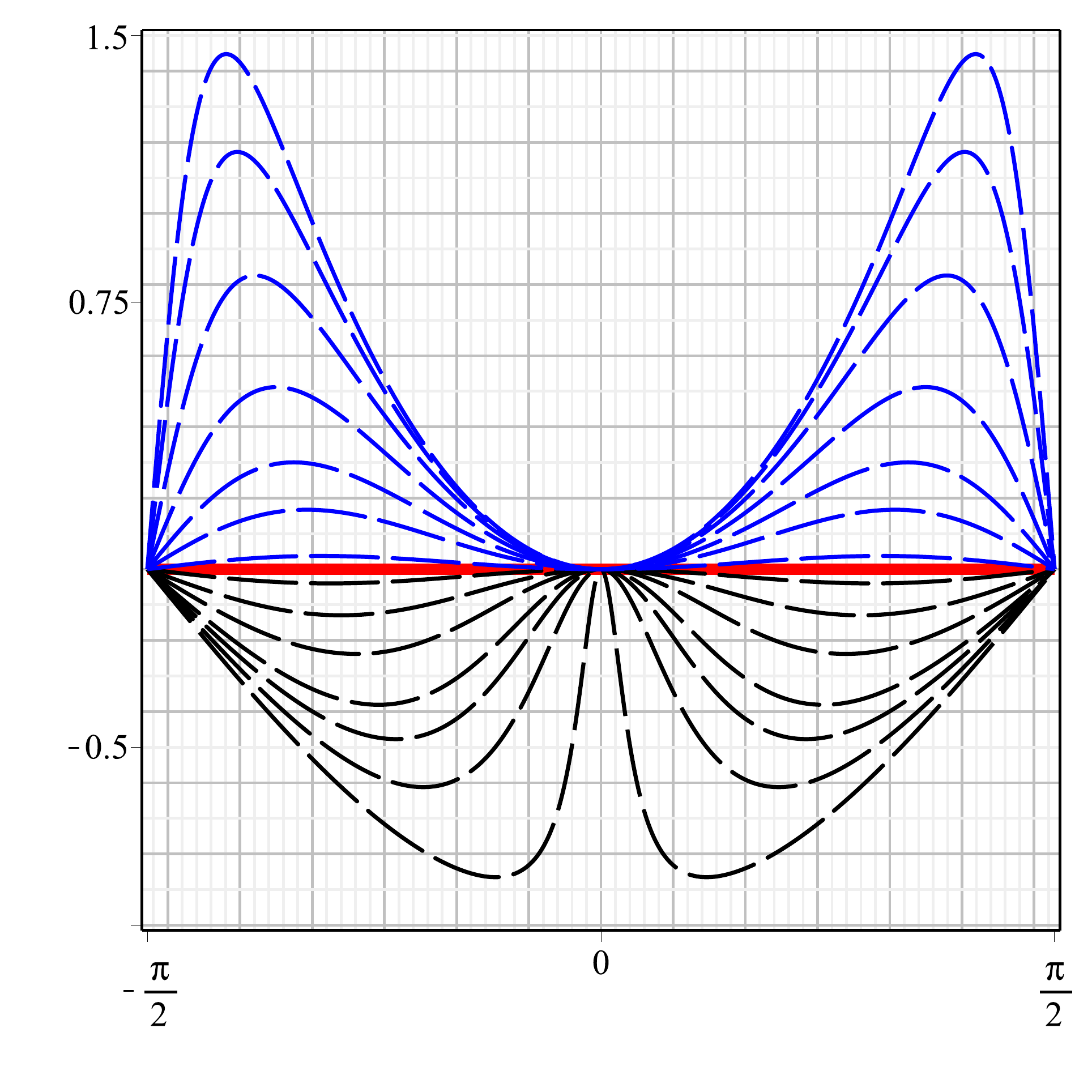}
\caption{The function $f(\phi)$ in Eq.~\eqref{f_a_sine} for $0<a\leq1$ (black) and for $a>1$ (blue) for several values of $a$. The red line represents the limit $a=1$ that is the standard case.}
\label{figfsine}
\end{figure}

Following the direction of Sec.~\ref{ex2}, we take a function $f(\phi)$ that depends on the parameter $a$ in the form
\be\label{f_a_sine}
f(\phi)=\frac{(a-1)\sin^2(\phi)\cos(\phi)}{1+(a-1)\cos^2(\phi)}.
\ee
The valid range for $a$ is found from the NEC in Eq.~\eqref{NEC}, which takes the form
\be
\frac{a}{1+(a-1)\cos^2(\phi)}\geq 0.
\ee
Since $\cos^2(\phi)\geq0$, the above expression is true only for $a\geq0$. We will again ignore $a=0$ because it leads to null energy density. For $a=1$, it recovers the standard case $f_1(\phi)=0$ and for $a\to\infty$ we get $f_\infty(\phi)=\sin(\phi)\tan(\phi)$. The function $f(\phi)$ in Eq.~\eqref{f_a_sine} is shown in Fig.~\ref{figfsine} for several values of $a$. We can calculate the energy density from \eqref{rho}, which becomes
%%%%%%%%%%%%%%%%%%%%
%%%%%%%%%%%%%%%%%%%
\be\label{rho_sine}
T_{00}=\frac{a\cos^2(\phi)}{1+(a-1)\cos^2(\phi)}=\frac{a\,S^2(x)}{1+(a-1)\,S^2(x)}.
\ee	
%%%%%%%%%%%%%%%%%%%
\begin{figure}[!htb]
\centering
\includegraphics[height=5cm]{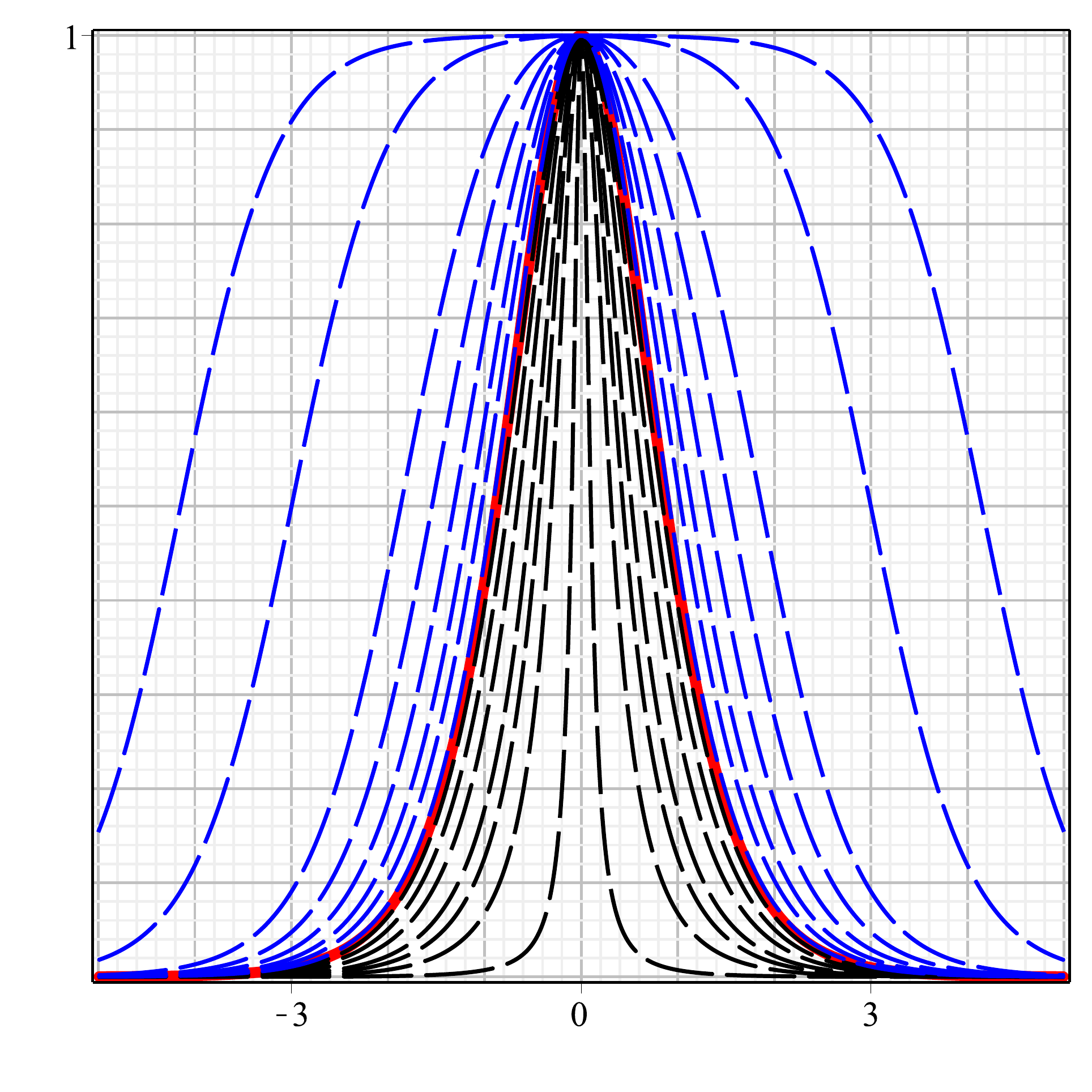}
\includegraphics[height=5cm]{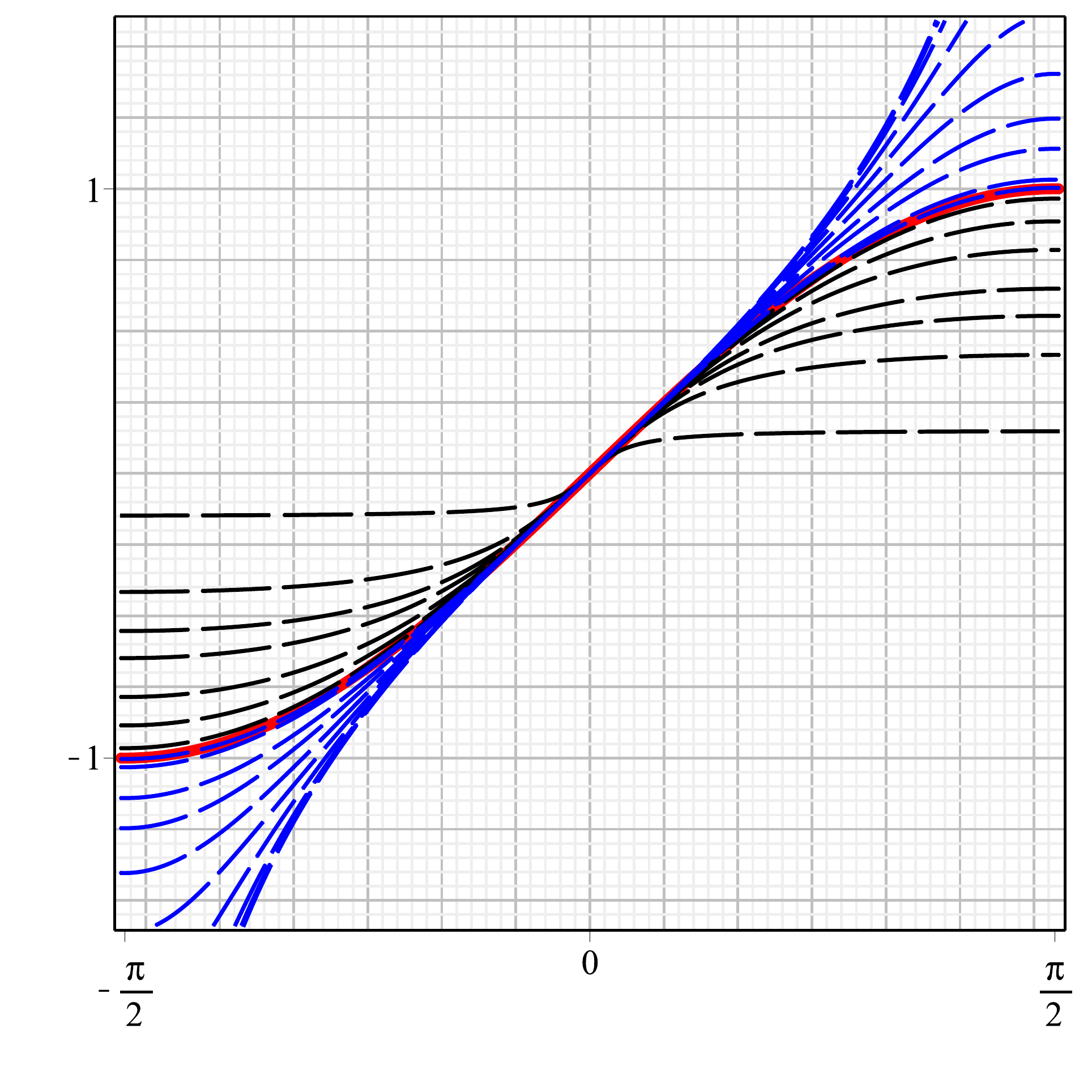}
\caption{The energy density in \eqref{rho_sine} as a function of $x$ (left) and the function $W(\phi)$ in Eq.~\eqref{W_sine} (right) for several values of $a$. The colors of the lines follow the previous figure.}
\label{figwsine}
\end{figure} 
%%%%%%%%%%%%%%%%%%%%
In order to find the function $W(\phi)$, we use Eq.~\eqref{Wp} and integrate it to get
\be\label{W_sine}
W(\phi)=\frac{\sqrt{a}\,M(\phi)}{|a-1|^{\frac12}},
 \ee
with
\be
M(\phi)=
\begin{cases}
\displaystyle\arctan\left(\sqrt{\frac{1-a}{a}}\,\sin(\phi)\right)
,&a\leq1\\
\displaystyle\arctanh\left(\sqrt{\frac{a-1}{a}}\,\sin(\phi)\right), & a>1.
\end{cases}
\ee
Given the general form of the function $W(\phi)$, we have $W(\phi)=\sin(\phi)$ for $a=1$ and $W(\phi)=\arctanh(\sin(\phi))$ for $a\to\infty$. The energy density in Eq.~\eqref{rho_sine} and the function $W(\phi)$ in Eq.~\eqref{W_sine} are plotted in Fig.~\ref{figwsine}. The energy can be calculated straightforwardly using Eq.~\eqref{ew}. For $a\leq1$, the energy is $E=2\arctan(\sqrt{(1-a)/a})\sqrt{a}/|a-1|^{1/2}$ and for $a>1$ it is $E=2\,\arctanh(\sqrt{(a-1)/a})\sqrt{a}/|a-1|^{1/2}$. In the limit $a\to1$, we recover $E=2$ as in the standard case and for $a\to\infty$, we get infinite energy, so we avoid this limit here.

\begin{figure}[htb!]
\centering
\includegraphics[width=5cm]{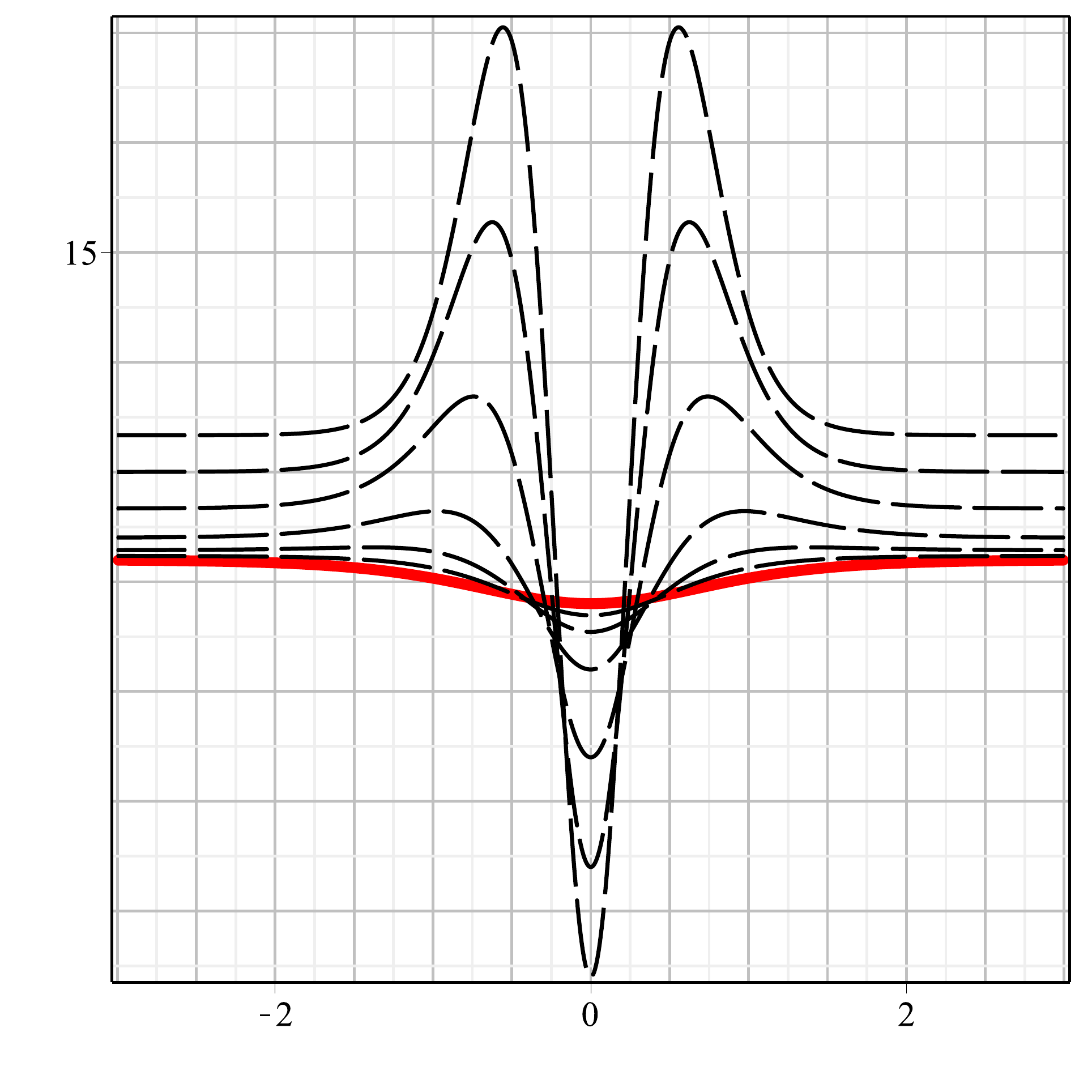}
\includegraphics[width=5cm]{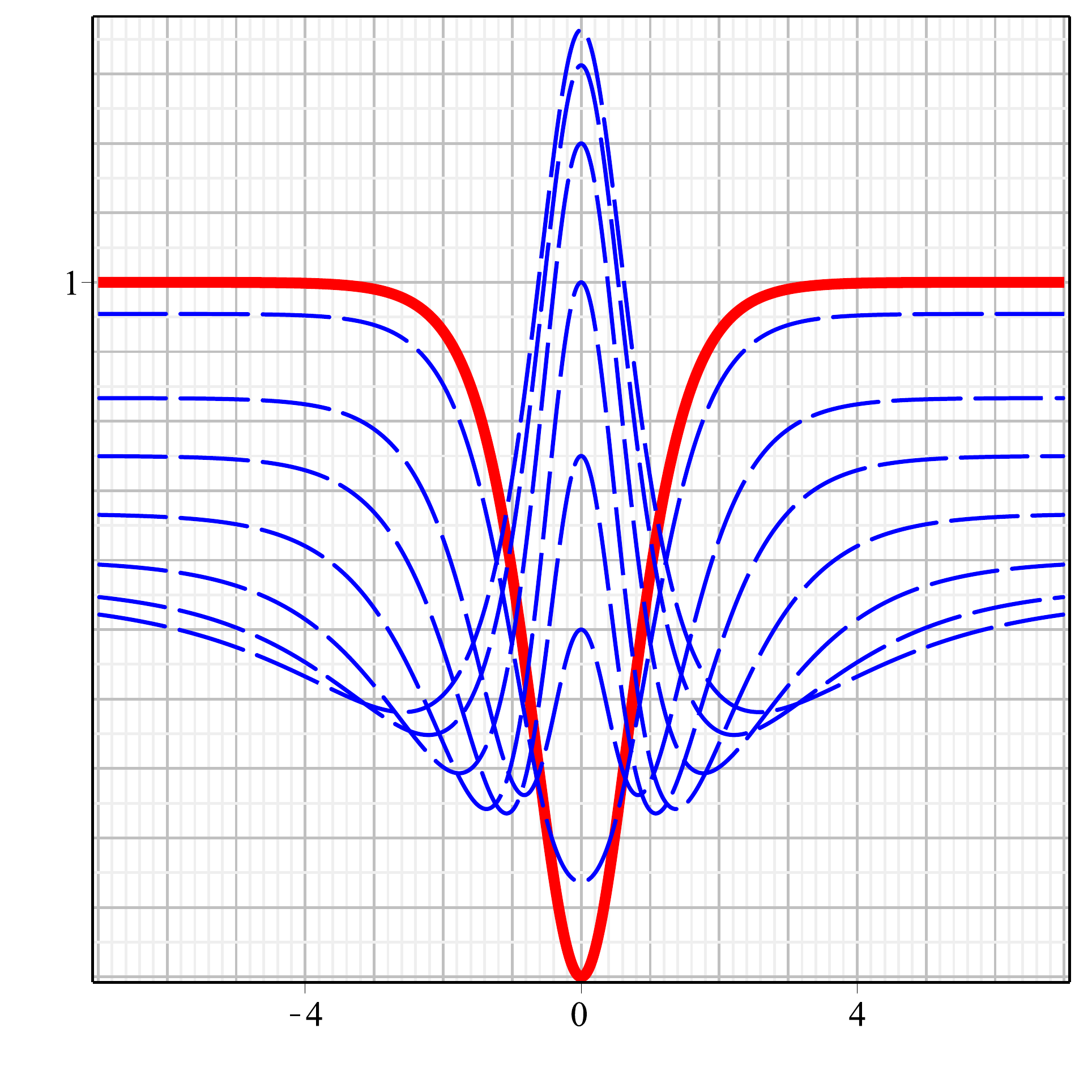}
\caption{The stability potential $U(z)$ in Eq.~\eqref{Uest_sine} for $0<a\leq1$ (black) and for $a>1$ (blue) for several values of $a$. The red line represents the limit $a=1$ that is the standard case.}
\label{figusine}
\end{figure}

The function $A$ in Eq.~\eqref{A} is such that
\be\label{A2_sine}
A^2=\frac1a+\frac{a-1}{a}\cos^2(\phi)=\frac1a+\frac{a-1}{a}S^2(x).
\ee
Notice that even though the expression for $A^2$ in terms of $\phi$ is different from the one in Eq.~\eqref{A2}, it is equal when written explicitly on $x$. This is an interesting feature, because it allows for the analytical correspondence between the variables $x$ and $z$ exactly in the form presented in Eq.~\eqref{zx2}. Nevertheless, the functions $g(z)$ and $U(z)$ are different. In this case, we obtain the function $g(z)$ from Eq.~\eqref{ssup}, which is explicitly given by
\begin{equation}
g(z)=-\tanh(z/\sqrt{a})\frac{\left(3(a-1)S^2(z/\sqrt{a})-a\right)}{\sqrt{a}\left((a-1)S^2(z/\sqrt{a})-a\right)}.
\end{equation}
The stability potencial in Eq.~\eqref{potestschr} takes the form
\begin{equation}\label{Uest_sine}
	U(z)=-\frac{12(a-1)S^4(z/\sqrt{a}) - (11a-9)S^2(z/\sqrt{a}) + a}{a\left((a-1)S^2(z/\sqrt{a})-a\right)}.
\end{equation}
It is plotted for several values of $a$ in Fig.~\ref{figusine}. Notice that despite the change in the potential $V(\phi)$, we were able to find a family of functions $f(\phi)$, guided by the parameter $a$, that allows for the presence of the analytical quantities in the study of stability. Here, however, the limit $a\to\infty$ leads to infinite energy, so we avoid it.
%%%%%%%%%%%%%%%%%%%%%%%%%%%
%%%%%%%%%%%%%%%%%%%%%%%%%%%%%%%%%%%%

\section{Braneworld Scenario}\label{sec3}
We now investigate the model in the braneworld scenario with a single dimension of infinite extent. The action is
\be\label{sbrane}
S=\int  d^4xdy \sqrt{g}\left(-\frac14 R  + X + \frac{2f(\phi)X}{\sqrt{|2X|}} -V(\phi)\right),
\ee
where $y$ denotes the extra dimension. As one knows, the metric tensor is
\be
g_{\mu\nu} = {\rm diag} \left(e^{2A(y)}, -e^{2A(y)}, -e^{2A(y)}, -e^{2A(y)}, -1 \right),
\ee
where $A(y)$ is the warp function and $e^{2A}$ is the warp factor. We also consider the scalar field depending only on the extra dimension $y$. Since we are interested in kinklike solutions, which are monotonic, for the scalar field we take $\phi^\prime>0$ and write the equation of motion as
\be\label{eomstaticBRANE}
\phi^{\prime\prime} + 4\phi^\prime A^\prime + f(\phi) A^\prime = V_\phi.
\ee
In the above equation, the prime denotes the derivative with respect to $y$. Notice that, unlike the flat spacetime equation of motion, the modification introduced by the cuscuton term, which is controlled by the function $f(\phi)$ appears here in the geometric contribution driven by $A^\prime$. Similarly to the flat case, though, it does not modify the term of second order. The Einstein equations become
\bes\label{einstein}
\begin{align}
A^{\prime\prime}  +\frac23\left(\phi^\prime + f(\phi)\right)\phi^{\prime} &=0,\\
{A^{\prime}}^2 - \frac16{\phi^\prime}^2 +\frac13V(\phi) &=0.
\end{align}
\ees
We then see that the equations which govern the system are of second order and present couplings between the scalar field and the warp function. In order to get a first order framework, we follow Ref.~\cite{trilogy3} and introduce an arbitrary function $W(\phi)$ in a manner that the potential can be written in the form
\be\label{potbrana}
V(\phi)=\frac18\left(W_\phi-2f(\phi)\right)^2-\frac13 W^2(\phi).
\ee
By doing so, we get the first order equations
\bes
\begin{align}\label{abrane}
	A^\prime &=-\frac13 W(\phi),\\ \label{fophibrane}
\phi^\prime &=\frac12 W_\phi-f(\phi),
\end{align}
\ees
which solves Eqs.~\eqref{eomstaticBRANE} and \eqref{einstein}. In order to get solutions that connects the minima of the potential, we require the function $f(\phi)$ to vanish in these points. The example considered in Sec.~\ref{ex1}, for instance, would fulfill this condition for $f_0=0$, which is the well-known standard model. Nevertheless, we may consider the case engendered by the function in Eq.~\eqref{f_a_sinh}, which satisfies the aforementioned condition. For instance, if one considers $W_{\textrm{brane}}(\phi) = 2W_\textrm{flat}(\phi)$, with $W_\textrm{flat}(\phi)$ given by Eq.~\eqref{W2}, the first order equation \eqref{fophibrane} admits exactly the same solution of Eq.~\eqref{tanh} with the change $x\to y$.

\begin{figure}[t!]
\centering
\includegraphics[width=5cm]{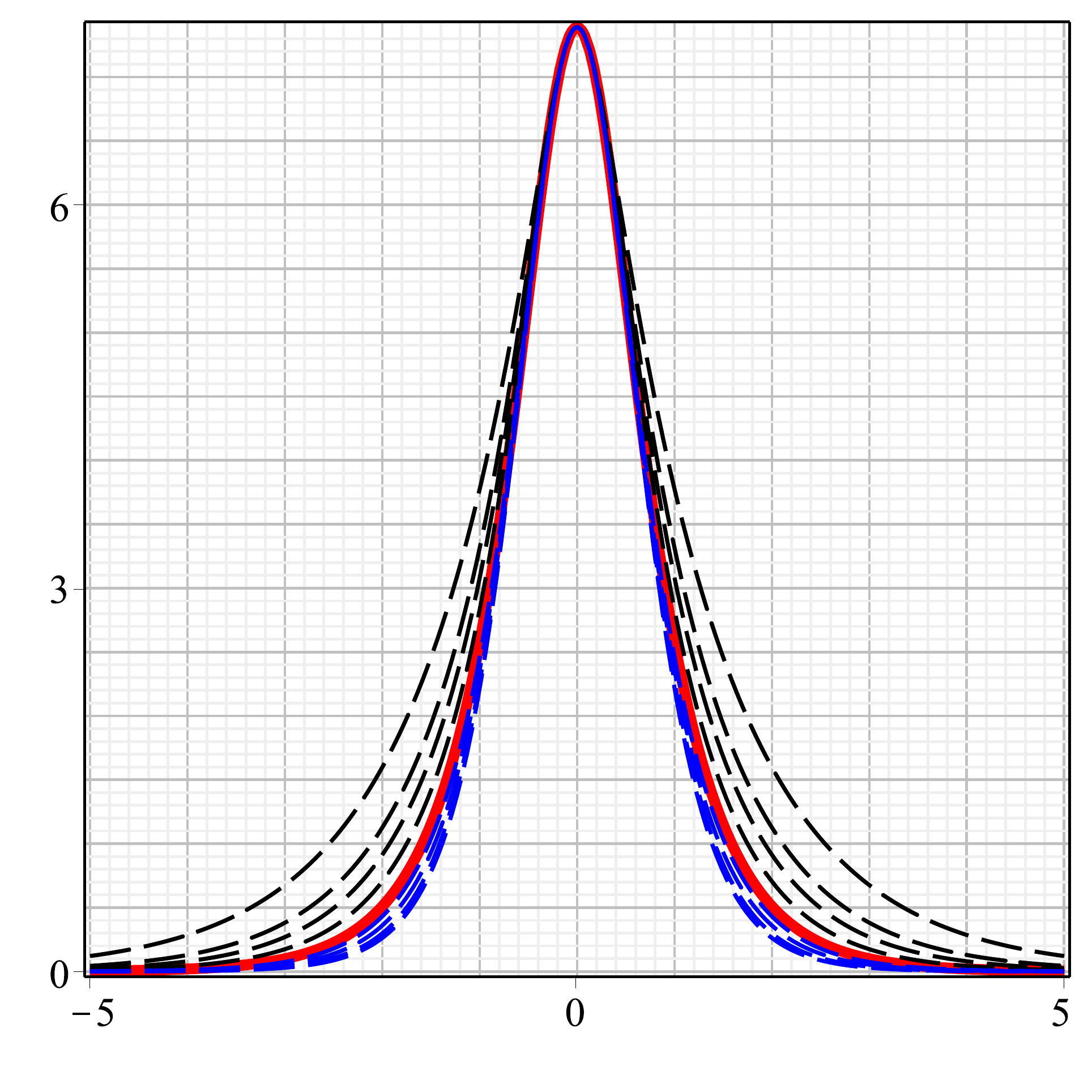}
\includegraphics[width=5cm]{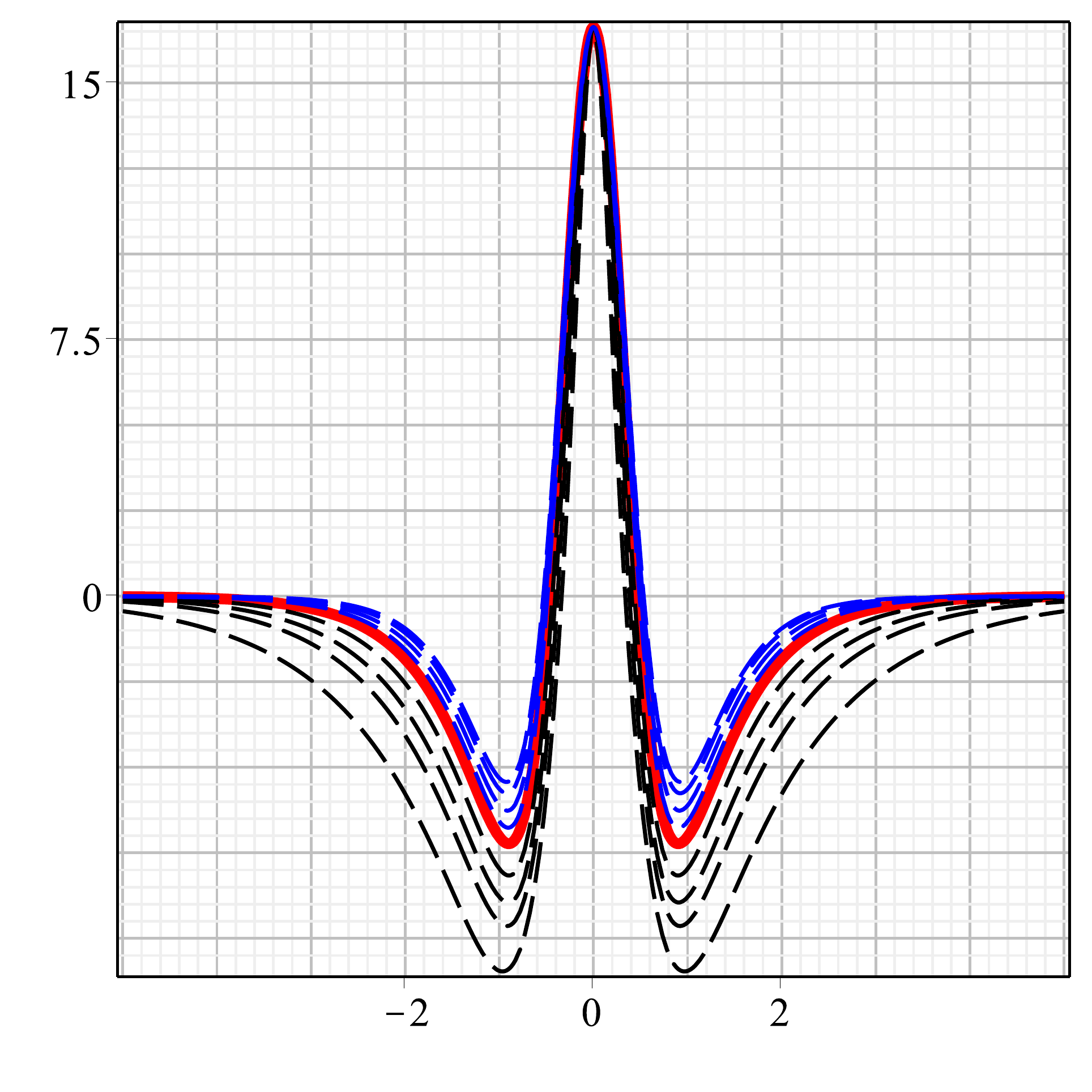}
\caption{The warp factor whose warp function is the solution of Eq.~\eqref{abrane} (left) and the energy density in Eq.~\eqref{rhobrane} (right) for several values of $a$. The colors are taken as in the previous figures.}
\label{figarhobrana}
\end{figure}

We, however, consider another model, defined by the functions
\bes\label{branemodel}
\begin{align}
f(\phi) &= \frac{(a-1)\phi^2(3 - 2\phi^2)}{3a - 2(a-1)\phi^2}, \\
W(\phi) &= \frac{3a\phi}{a-1} + \frac32\frac{\sqrt{6a}\, M_b(\phi)}{|a-1|^{3/2}},
\end{align}
\ees
where
\be
M_b(\phi)=
\begin{cases}
\displaystyle\arctan\left(\sqrt{\frac{2(1-a)}{3a}}\,\phi\right)
,&a\leq1\\
-\displaystyle\arctanh\left(\sqrt{\frac{2(a-1)}{3a}}\,\phi\right), & a>1.
\end{cases}
\ee
It is motivated by the possibility of having analytical expressions in the study of the brane stability. In this case, the solution of the first order equation \eqref{fophibrane} is
\be
\phi(y) = \frac{\sqrt{6}}{2}\tanh\left(\frac{\sqrt{6}}{2}\,y\right).
\ee
This solution connects $\phi=-\sqrt{6}/2$ to $\phi=\sqrt{6}/2$ asymptotically. The potential can be calculated from Eq.~\eqref{potbrana}, which leads to
\be\label{potb1}
V(\phi)=\frac12\left(3-2\phi^2\right)^2-\frac13 W^2(\phi).
\ee
%%%%%%%%%%%%%%%%%%%%%%%%%%%
\begin{figure}[t!]
\centering
\includegraphics[width=6cm]{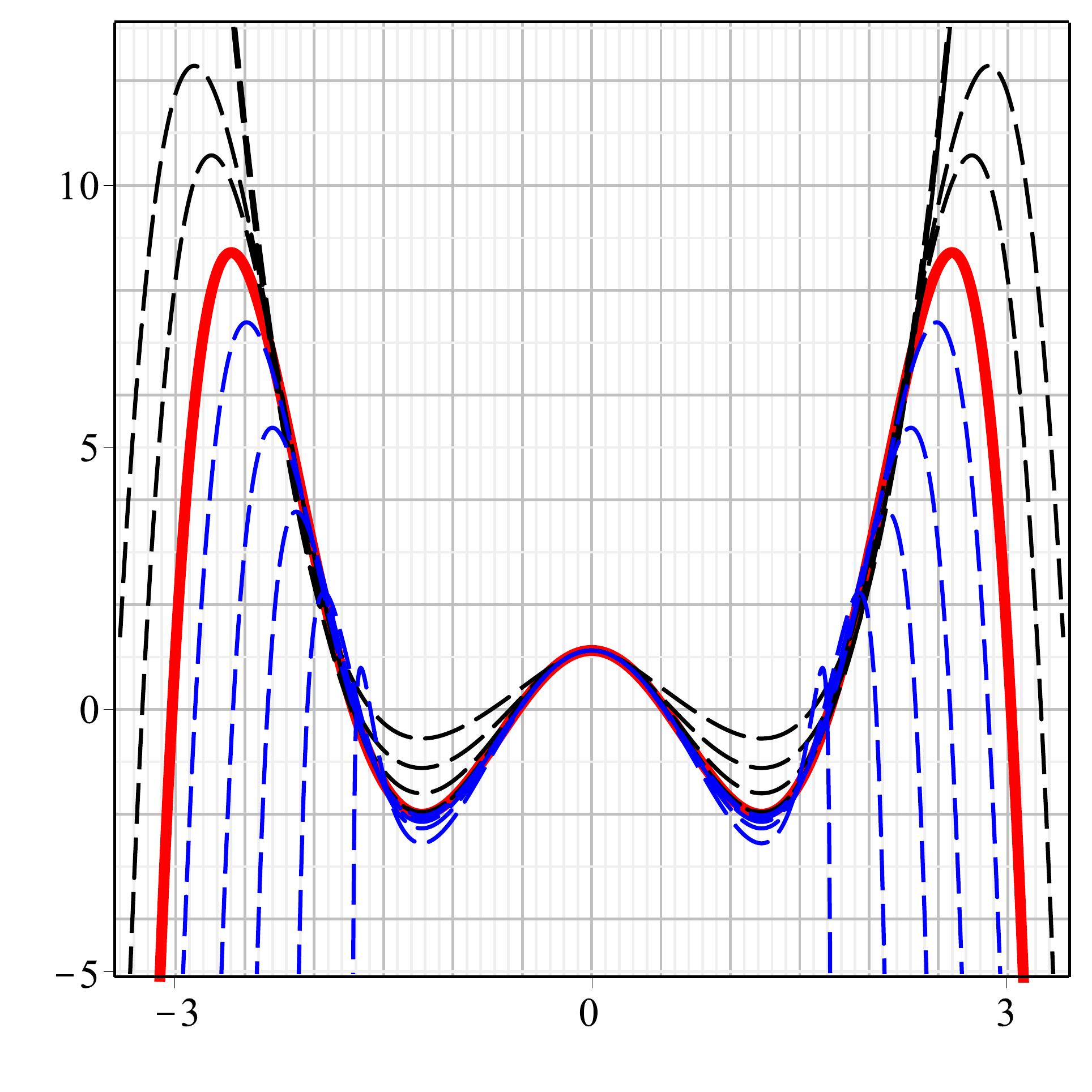}
\caption{The potential in Eq.~\eqref{potb1} for several values of $a$. The colors are taken as in the previous figures.}
\label{figpotbrana}
\end{figure}
%%%%%%%%%%%%%%%%%%%%%%%%%%%
For $a=1$, we have $V(\phi) = (3-2\phi^2)^2/2 - \phi^2(9-2\phi^2)^2/27$, which is the standard case, $f(\phi)=0$. On the other hand, the limit $a\to\infty$ leads to $V(\phi) = (3-2\phi^2)^2/2 - 3\phi^2$. In Fig.~\ref{figpotbrana}, we plot this potential for several values of $a$. Its minima are located at $\phi=\pm\sqrt{6}/2$, with a local maximum between them at $\phi=0$. The minima goes deeper as $a$ increases, whilst the maximum remains unchanged. Regarding the warp function, we must solve Eq.~\eqref{abrane}. Even though an analytical solution may be found, it is cumbersome, so we omit it here. For $a=1$, it simplifies to $A(y) = 1-2\ln(\cosh(\sqrt{6}\,y/2))/3 - \tanh^2(\sqrt{6}\,y/2)/6$.  The limit $a\to\infty$ leads to $A(y) = 1 - \ln(\cosh(\sqrt{6}\,y/2))$. In Fig.~\ref{figarhobrana}, we plot it for several values of $a$.

We may follow Ref.~\cite{trilogy3} and calculate the energy density. It is
\be\label{rhobrane}
\rho_\textrm{brane}(y) = e^{2A(y)}\left(\frac12{\phi^\prime}^2 + \phi^\prime f(\phi) + V(\phi)\right).
\ee
The explicit expression for a general $a$ is cumbersome. For $a=1$ it is $\rho(y) = \exp(2A(y)) (2S^6(y) + 15 S^4(y) - 8)$. In the limit $a\to\infty$ it simplifies to $\rho(y) = 9e^2(3S(y)^2-2)S^2(y)/4$. For simplicity, we have adopted the notation $S(y) = \sech(\sqrt{6}\,y/2)$. In Fig.~\ref{figarhobrana}, we plot the above energy density for several values of $a$. We see that as $a$ increases, the width of both the warp factor and the energy density shrinks.
%%%%%%%%%%%%%%%%%%%%%%%%%%%
%%%%%%%%%%%%%%%%%%%%%%%%%%%

We then turn our attention to the brane stability.  it is important to know if the modification in the action \eqref{sbrane} due to the cuscuton term destabilizes the gravity sector of the brane. This issue is investigated with perturbations of the metric and of the field. We follow Ref.~\cite{trilogy3}, which lead us to the stability equation
\be\label{eqest}
\left(-\frac{\partial^2}{\partial z^2} + U(z)\right)H_{\mu\nu} = p^2 H_{\mu\nu},\quad\text{with}\quad dz = e^{-A(y)}dy.
\ee
Here, $H_{\mu\nu}$ represents an scaled perturbation. The stability potential, $U(z)$ is
\be\label{potest}
U(z) = \frac94A_z^2 +\frac32{A_{zz}},
\ee
with $A_z=\partial A/\partial z$. As it is known, the stability equation may be written as
\be
\left(\frac{\partial}{\partial z} + \frac32 {A_z}\right)\left(-\frac{\partial}{\partial z} + \frac32 {A_z}\right)H_{\mu\nu} = p^2 H_{\mu\nu},
\ee
which shows the absence of graviton bound states with negative mass. This ensures the stability of the brane.

In order to write the explicit expression for the stability potential \eqref{potest}, we must firstly find the conformal coordinate $z$ and its inverse. This is a very hard task to be performed analytically. In our model, defined by Eq.~\eqref{branemodel}, for instance, we have not been able to calculate it for a general $a$. Nonetheless, in the limit $a\to\infty$, we have found
\be\label{zybrane}
z=\frac{\sqrt{6}}{3e}\sinh\left(\frac{\sqrt{6}}{2}\,y\right).
\ee
This leads to the stability potential
\be
U(z) = \frac{9e^2}{4}\frac{15e^2z^2 - 4}{(3e^2z^2 + 2)^2}.
\ee
Again, we emphasize that the presence of the last two equations is very hard to be obtained in general. Moreover, differently from the flat spacetime case, even by choosing the function $f(\phi)$ properly, here we were only able to get the analytical expressions in the study of the linear stability for $a\to\infty$. The expression in Eq.~\eqref{zybrane} that relates $y$ and $z$, and the above stability potential are similar to the one for the sine-Gordon case investigated in Ref.~\cite{gremm} without the presence of the cuscuton term. In this sense, the function $f(\phi)$ that drives the cuscuton term is important in the non-sine-Gordon cases because it may allow for the presence of analytical quantities in the study of the brane stability.

\section{Outlook}\label{sec4}
We have investigated the presence of kinks in generalized models that adds the cuscuton term in the Lagrangian density, which has the form \eqref{lmodel}. We have seen that its associated equations of motion are the very same of the standard case for static configurations. The energy density and the linear stability, though, are modified by the new term. Even so, the solutions are stable and admit stability equations that are of the Sturm-Liouville type and can be mapped into ones of the Schr\"odinger type. It is interesting to point that we have been able to find model which is completely described by analytical quantities.

The investigation was extended to the curved spacetime, in the braneworld scenario with an extra dimension of an infinite extent. This case is different from the flat spacetime one: the cuscuton term appears in the geometric term of the scalar field equation of motion. In order to simplify the problem, we have found a first order formalism that is compatible with the Einstein and scalar field equations. Even though the curved spacetime scenario brings more difficulties than the flat one, we have been able to show the brane stability and to calculate its associated Schr\"odinger-like potential for an specific model, which is solely described by analytical expressions.

We highlight that a similar treatment can be conducted with the cuscuton term added to an unspecified generalized model. In the flat spacetime, one can look into the action
\be
S = \int dx\,dt \left(\LL(X,\phi) + \frac{2f(\phi)X}{\sqrt{|2X|}} \right),
\ee
where $\LL(X,\phi)$ is supposed to give rise to kinklike solutions. In this case, the equations of motion for static, unidimensional configurations are
\be
\left(\LL_X \phi^\prime\right)^\prime + \LL_\phi = 0.
\ee
Thus, it admits the very same field profile of an action with $f(\phi)=0$, which represents the absence of the cuscuton term in the model. The energy density, though, is given by $\rho = -\LL + f(\phi)|\phi^\prime|$. One may show that the stability is also modified by $f(\phi)$. 

An interesting direction for future research concerns the time dependence of these solutions. One may also investigate how the integrability of known models, such as the sine-Gordon, is modified by the inclusion of the cuscuton term driven by function $f(\phi)$. Another possibility is to investigate how this term affects other topological structures such as vortices and monopoles. Some of these issues are currently under consideration and we hope to report them elsewhere.

\acknowledgements{We would like to acknowledge the Brazilian agency CNPq for partial financial support. IA thanks support from grant 140490/2018-3, MAM thanks support from grant 155551/2018-3 and RM thanks support from grant 306826/2015-1.}


\begin{thebibliography}{99}
\bibitem{vilenkin} A. Vilenkin and E.P.S. Shellard, \textit{Cosmic Strings and Other Topological Defects}, Cambridge University Press (2000).
\bibitem{manton} N. Manton and P. Sutcliffe, \textit{Topological Solitons}, Cambridge University Press (2004).
\bibitem{weinberg} E.J. Weinberg, \textit{Classical solutions in Quantum Field Theory: Solitons and Instantons in High Energy Physics}, Cambridge University Press (2012).
\bibitem{vachaspati} T. Vachaspati, \textit{Kinks and Domain Walls: An Introduction to Classical and Quantum Solitons}, Cambridge University Press (2006).
\bibitem{fradkin} E. Fradkin, \textit{Field Theories of Condensed Matter Physics}, Cambridge University Press (2013).

\bibitem{cosm1} C. Armendariz-Picon, T. Damour, and V. Mukhanov, Phys. Lett. B {\bf458}, 209 (1999).
\bibitem{cosm2} C. Armendariz-Picon, V. Mukhanov, and P.J. Steinhard, Phys. Rev. Lett. {\bf85}, 4438 (2000). 
\bibitem{cosm3} C. Armendariz-Picon, V. Mukhanov, and P.J. Steinhardt, Phys. Rev. D {\bf63}, 103510 (2001).
\bibitem{jackiwcs} R. Jackiw and E.J. Weinberg, %``Selfdual Chern-simons Vortices,''
Phys. Rev. Lett.  {\bf 64}, 2234 (1990).
\bibitem{G1}J. Lee and S. Nam, Phys. Lett. B {\bf261}, 437 (1991).
\bibitem{G2}D. Bazeia, Phys. Rev. D {\bf46}, 1879 (1992).
\bibitem{godvortex} D. Bazeia, L. Losano, M.A. Marques, R. Menezes, and I. Zafalan, Nucl. Phys. B {\bf 934}, 212 (2018).
\bibitem{compvortex} D. Bazeia, L. Losano, M.A. Marques, R. Menezes, and I. Zafalan, %\textit{Compact vortices},
Eur. Phys. J. C {\bf 77}, 63 (2017).
\bibitem{compcs} D. Bazeia, L. Losano, M.A. Marques, and R. Menezes, Phys. Lett. B {\bf772}, 253 (2017).

\bibitem{stabbabichev} E. Babichev, %Global topological k-defects,''
Phys. Rev. D {\bf 74}, 085004 (2006).
\bibitem{trilogy1} D. Bazeia, L. Losano, R. Menezes, and J.C.R.E. Oliveira, Eur. Phys. J. C {\bf 51}, 953 (2007).
\bibitem{trilogy2} D.Bazeia, L.Losano, and R. Menezes, %\textit{First-order framework and generalized global defect solutions},
Phys. Lett. B {\bf668}, 246 (2008).
\bibitem{trodden}M. Andrews, M. Lewandowski, M. Trodden, and D. Wesley, Phys. Rev. D {\bf82}, 105006 (2010).
\bibitem{twinb1}D. Bazeia, J.D. Dantas, A.R. Gomes, L. Losano, and R. Menezes, Phys. Rev. D {\bf84}, 045010 (2011).
\bibitem{twinb2}D. Bazeia and R. Menezes, Phys. Rev. D {\bf84}, 125018 (2011).
\bibitem{twinb3}D. Bazeia, A.S. Lobao, Jr., and R. Menezes, Phys. Rev. D {\bf86}, 125021 (2012).
\bibitem{twinb4}D. Bazeia, A.S. Lobao, L. Losano, and R. Menezes, Eur. Phys. J. C {\bf74}, 2755 (2014).
\bibitem{twinb5} A.R. Gomes, R. Menezes, K.Z. Nobrega, and F.C. Simas, Phys. Rev. D {\bf90}, 065022 (2014).
\bibitem{twinb6} D. Bazeia, M.A. Marques, and R. Menezes, %``Twinlike Models for Kinks, Vortices and Monopoles,''
Phys. Rev. D {\bf 96}, 025010 (2017).
\bibitem{bil} D. Bazeia, M.A. Marques, and R. Menezes, %``Generalized Born-Infeld–like models for kinks and branes,''
EPL {\bf 118}, 11001 (2017).
\bibitem{fphil} L.Losano, M.A. Marques, and R.Menezes, %``Generalized scalar field models with the same energy density and linear stability,''
Phys. Lett. B {\bf 775}, 178 (2017).
\bibitem{S}A. Sen, JHEP {\bf9912}, 027 (1999). 
\bibitem{S0}A. Sen and B. Zwiebach, JHEP {\bf0003}, 002 (2000).
\bibitem{S1}A. Sen, Phys. Rev. D {\bf68}, 066008 (2003).
\bibitem{T1}G.W. Gibbons, K. Hashimoto, and S. Hirano, JHEP {\bf0909}, 100 (2009). 
\bibitem{T2}F.A. Brito, H.S. Jesuino, JHEP {\bf1007}, 031 (2010).
\bibitem{D}N. Callebaut and D. Dudal, JHEP {\bf1401}, 055 (2014).
\bibitem{T3}M. Saidy-Sarjoubi and  D. Kamani, Phys. Rev. D {\bf92}, 046003 (2015). 
\bibitem{pad}T. Padmanabhan, Phys. Rev. D {\bf66}, 021301 (2002).
\bibitem{baz}D. Bazeia, C.B. Gomes, L. Losano, and R. Menezes, Phys. Lett. B {\bf633}, 415 (2006). 
\bibitem{berto}A.E. Bernardini and O. Bertolami, Phys. Lett. B {\bf726}, 512 (2013). 
\bibitem{gen1} X. Jin, X. Li, and D. Liu, Class. Quantum Grav. {\bf24}, 2773 (2007).
\bibitem{gen2} S. Sarangi, %``DBI global strings,''
JHEP {\bf 0807}, 018 (2008).
\bibitem{gen3} R. Casana, M.M. Ferreira, Jr, and E. da Hora, Phys. Rev. D {\bf86}, 085034 (2012).
\bibitem{gen4} H.S. Ramadhan, Phys. Lett. B {\bf758}, 149 (2016).
\bibitem{gen5} C. Adam and J.M. Queiruga, Phys. Rev. D {\bf85}, 025019 (2012).
\bibitem{gen6} Y. Zhong and Y.-X. Liu, Class. Quant. Grav. {\bf32}, 165002 (2015).
\bibitem{gen7} C. Adam, J.M. Queiruga, Phys. Rev. D {\bf84}, 105028 (2011).

\bibitem{cuscuton1} N. Afshordi, D. J. Chung, and G. Geshnizjani, %Cuscuton: A Causal Field Theory with an Infinite Speed of Sound,
Phys. Rev. D {\bf75}, 083513 (2007).
\bibitem{cuscuton2} N. Afshordi, D. J. Chung, M. Doran, and G. Geshnizjani, %Cuscuton Cosmology: Dark Energy meets Modified Gravity,
Phys. Rev. D {\bf75}, 123509 (2007).
\bibitem{cuscuton3}J. Magueijo, %``Speedy sound and cosmic structure,''
Phys. Rev. Lett.  {\bf 100}, 231302 (2008).
\bibitem{cuscuton4} N. Afshordi, %``Cuscuton and low energy limit of Horava-Lifshitz gravity,''
Phys. Rev. D {\bf 80}, 081502 (2009).
\bibitem{cuscuton5} M. Li, X.D. Li, S. Wang, and Y. Wang, %``Dark Energy,''
Commun. Theor. Phys.  {\bf 56}, 525 (2011).
\bibitem{cuscuton6} D. Bazeia, F.A. Brito, and F.G. Costa, %``Braneworld solutions from scalar field in bimetric theory,''
Phys. Rev. D {\bf 87}, 065007 (2013).
\bibitem{cuscuton7} H. Gomes and D.C. Guariento, %``Hamiltonian analysis of the cuscuton,''
Phys. Rev. D {\bf 95},104049 (2017).
\bibitem{cuscuton8} S.S. Boruah, H.J. Kim, and G. Geshnizjani, %``Theory of Cosmological Perturbations with Cuscuton,''
JCAP {\bf 1707}, 022 (2017).

\bibitem{RS} L. Randall and R. Sundrum, Phys. Rev. Lett. {\bf83}, 4690 (1999).
\bibitem{GW} W.D. Goldberger and M.B. Wise, Phys. Rev. Lett. {\bf83}, 4922, (1998).
\bibitem{MC} M. Cvetic, S. Griffies, and S.-J. Rey, Nucl. Phys. B {\bf381}, 301 (1992).
\bibitem{brane} O. DeWolfe, D.Z. Freedman, S. Gubser, and A. Karch, Phys. Rev. D {\bf62}, 046008 (2000).
\bibitem{cvetic} M. Cvetic and N.D. Lambert, Phys. Lett. B {\bf 540}, 301 (2002).
\bibitem{gremm} M. Gremm, Phys. Lett. B {\bf 478}, 434 (2000).
\bibitem{fktc} D. Bazeia, L. Losano, M.A. Marques, and R. Menezes, %``From Kinks to Compactons,''
Phys. Lett. B {\bf 736}, 515 (2014).
\bibitem{fkthc} D. Bazeia, M.A. Marques, and R. Menezes, %``Models for asymmetric hybrid brane,''
Phys. Rev. D {\bf 92}, 084058 (2015).
\bibitem{trilogy3} D. Bazeia, A.R. Gomes, L. Losano, and R. Menezes, %``Braneworld Models of Scalar Fields with Generalized Dynamics,''
Phys. Lett. B {\bf 671}, 402 (2009).

\end{thebibliography}
\end{document}